\documentclass[prx,twocolumn,superscriptaddress,nofootinbib,showpacs,notitlepage]{revtex4-2}
\usepackage{amsmath}
\usepackage{amsfonts}
\usepackage{amssymb}
\usepackage{bbm}
\usepackage[utf8]{inputenc}
\usepackage{braket}
\usepackage{environ}
\usepackage{graphicx,color,dsfont}
\usepackage{booktabs}
\usepackage{floatrow}
\usepackage[caption=false,label font={bf,normalsize}]{subfig}
\usepackage[dvipsnames]{xcolor}
\usepackage[all]{xy}
\usepackage{relsize}
\usepackage{hyperref}
\usepackage{accents}
\hypersetup{linktocpage}
\definecolor{darkblue}{RGB}{0,0,127} 
\definecolor{darkgreen}{RGB}{0,150,0}
\hypersetup{colorlinks=true,citecolor=darkblue,linkcolor=darkblue, 
	filecolor=red, urlcolor=darkblue,pdftitle={Non-Abelian Self-Correcting Quantum Memory},
pdfauthor={Po-Shen Hsin, Ryohei Kobayashi, Guanyu Zhu}}
\usepackage{lmodern}

\usepackage{lipsum}

\makeatletter
\def\@opargbegintheorem#1#2#3{\trivlist
   \item[]{\bfseries #1\ #2\ (#3)} \itshape}
\makeatother

\NewEnviron{eqs}{%
\begin{equation}\begin{split}
    \BODY
\end{split}\end{equation}
}

\newcommand{\ZZ}{\mathbb{Z}}
\newcommand{\Z}{\mathbb{Z}}
\begin{document}

\title{Non-Abelian Self-Correcting Quantum Memory and \\ Transversal Non-Clifford Gate beyond the $n^{1/3}$ Distance Barrier }

\author{Po-Shen Hsin}
\email{po-shen.hsin@kcl.ac.uk}
\affiliation{Mani L. Bhaumik Institute for Theoretical Physics,
475 Portola Plaza, Los Angeles, CA 90095, USA}

\affiliation{Department of Mathematics, King’s College London, Strand, London WC2R 2LS, UK}

\author{Ryohei Kobayashi}
\email{ryok@ias.edu}
\affiliation{Department of Physics, Condensed Matter Theory Center, and Joint Quantum Institute, University of Maryland, College Park, Maryland 20742, USA}

\affiliation{School of Natural Sciences, Institute for Advanced Study, Princeton, NJ 08540, USA}

\author{Guanyu Zhu}
\email{guanyu.zhu@ibm.com}
\affiliation{IBM Quantum, IBM T.J. Watson Research Center, Yorktown Heights, NY 10598 USA}

\begin{abstract}
We construct a family of infinitely many new candidate non-Abelian self-correcting topological quantum memories in $D\geq 5+1$ spacetime dimensions without particle excitations using local commuting non-Pauli stabilizer lattice models and field theories of $\mathbb{Z}_2^3$ higher-form gauge fields with nontrivial topological action.
We call such non-Pauli stabilizer models magic stabilizer codes.
The family of topological orders have Abelian electric excitations and non-Abelian magnetic excitations that obey Ising-like fusion rules and non-Abelian braiding, including Borromean ring type braiding which is a signature of non-Abelian topological order, generalizing the dihedral group $\mathbb{D}_8$ gauge theory in (2+1)D. The simplest example includes a new non-Abelian self-correcting memory in (5+1)D with Abelian loop excitations and non-Abelian membrane excitations. We prove the self-correction property and the thermal stability, and devise a probabilistic local cellular-automaton decoder. We also construct fault-tolerant non-Clifford CCZ logical gate using constant depth circuit from higher cup products in the 5D non-Abelian code. The use of higher-cup products and non-Pauli stabilizers allows us to get an $O(n^{2/5})$ distance  overcoming the $O(n^{1/3})$ distance barrier in conventional topological stabilizer codes, including the 3D color code and the 6D self-correcting color code.

\end{abstract}

\maketitle
\tableofcontents

\section{Introduction}

Self-correcting quantum memories are highly desirable for fault-tolerant quantum computing due to its passive protection at the hardware level and the connection to a local cellular automaton decoder \cite{Dennis:2001nw, Brown_2016}, as well as single-shot error correction \cite{Bombin:2015hia}.    All previous known examples of self-correction are essentially toric codes in four spatial dimensions or higher~\cite{Dennis:2001nw,alicki2010thermal, Bombin:2013cv}, which are Abelian topological order and belong to the class of Pauli stabilizer models.   Recently, such 4D self-correcting model have also been realized experimentally on the ion-trap platform and shows the single-shot decoding advantage over 2D surface codes \cite{berthusen2024experiments4dsurfacecode}\footnote{
Higher dimensional topological orders above the physical dimension can be realized using long-range connections and are relevant for practical fault-tolerant quantum computation \cite{Guth:2014cj, Breuckmann:2020_single-shot_hyperbolic, freedman:2020_manifold_from_code}, and they are ubiquitous in quantum low-density parity-check (LDPC) codes \cite{Bravyi:2024wc, Breuckmann:2017hy, fiberbundlecode21,9567703,9490244,hastingswr21,pkldpc22,lh22,guefficient22,dhlv23,lzdecoding23,gusingleshot23, zhu2023non}. 
}.  However, these self-correcting models are quite limited in terms of the computation power.  For example, 4D loop toric code can only realize logical Clifford gates, which hence cannot be universal.  In order to get non-Clifford gates, one has to use 6D toric codes or color codes proposed by Bombin et al \cite{Bombin:2013cv}, which requires a space overhead $n=O(d^3)$, where $n$ represents the number of physical while $d$ is the code distance. On the other hand, since 2D non-Abelian topological code such as Fibonacci string-net model has exhibited a universal logical gate set \cite{Freedman_Larsen_wang_2002, levin2005, Koenig:2010do, schotte2020quantum, minev2024realizing}, a natural question is whether there exists a non-Abelian self-correcting memory at lower dimension which can potentially have higher computational power. 

From the perspective of quantum phases of matter, a fundamental question is that whether topological orders can exist at finite temperature.  It is known that 2D toric code is unstable at finite temperature \cite{Nussinov_2008,Bravyi_2009,PhysRevLett.107.210501,Brown_2016,Yoshida_2011} while the 4D loop toric code is stable \cite{Dennis:2001nw,alicki2010thermal, Bombin:2013cv}.   However, no examples beyond Abelian topological order or stabilizer models were known before this paper. This is because to find such non-Abelian self-correcting memory, one needs to have a topological order without particle excitations, which is quite challenging in the case of non-Abelian models.   In fact, a no-go theorem was shown for non-Abelian topological order without particle excitations in (4+1)D (space-time dimension) \cite{Johnson-Freyd:2020ivj,Johnson-Freyd:2021tbq,Cordova:2023bja}.   Moreover, the usual (1-form) non-Abelian discrete gauge theory (i.e., quantum double models) \cite{kitaev2003} will necessarily have particle-like excitations.

Our present work solves this issue by turning to the  higher-form gauge theory and in five or higher spatial dimensions.\footnote{
Higher-form gauge theories have been studied in e.g. \cite{Horowitz:1989ng,Witten:1998wy,Hitchin:2000jd,Maldacena:2001ss,Dijkgraaf:2004te,walker201131tqftstopologicalinsulators,Kapustin:2013uxa,Kapustin:2014gua,Hsin:2018vcg}, especially in high energy theory context. However, the properties of higher-form gauge theories with general gauge groups and general topological actions of the gauge fields in general dimensions remain unknown. In this work we study the properties of a family of particular higher-form gauge theories whose properties have not been explored.
}
In particular, we have developed a new type of topological quantum field theory (TQFT) based on a twisted version of $\mathbb{Z}_2^3$ higher-form gauge theory.  We further constructed the corresponding lattice models with non-Pauli (Clifford) stabilizers.  We dub such a model a \textit{magic stabilizer code} since it goes beyond the Pauli stabilizer formalism. We then prove that such non-Abelian models are indeed self-correcting, i.e., with an exponentially long memory time as system size grows, and hence is also thermally stable at finite temperature.   We have also developed a local cellular-automaton decoder for such models. In particular, the simplest non-Abelian self-correcting quantum memories we discover in five spatial dimensions  admit a single-shot non-Clifford logical CCZ gate via a constant-depth circuit with lower space-time overhead than the 3D color/surface codes \cite{Bombin:2015jk, Kubica:2015, vasmer2019} or the 6D self-correcting color code scheme \cite{Bombin:2013cv}.  In particular, the 5D non-Abelian code has distance $d=O(n^{2/5})$, which is beyond the $O(n^{1/3})$ distance barrier in conventional topological stabilizer code as implied by the Bravyi-K\"onig bound \cite{Bravyi:2013dx}, including the case of the 3D or 6D color/surface codes. 
Although another recent scheme in Ref.~\cite{zhu2025topological} achieves the $O(n^{1/2})$ distance by using the expansion properties of quantum LDPC codes, the present work considers a purely topological setting which can be realized with a small set of qubits (likely O(100)).  The distance enhancement in this case is mainly due to the use of higher-cup products in the logical gate construction and the fact that we utilize non-Pauli (Clifford) stabilizer models beyond the Pauli stabilizer codes.

Interestingly, we also find that our 5D non-Abelian self-correcting memory can also be obtained from a twisted compatification of the 6D self-correcting color code in Ref.~\cite{Bombin:2013cv} down to five spatial dimensions. 
The self-correcting quantum memories we discover have similar properties as the non-Abelian $\mathbb{D}_8$ topological order observed in another experiment by Quantinuum \cite{Iqbal:2023wvm}. Thus it is highly plausible that the new quantum memories can be realized by similar experiment setups on ion-trap platform using the long-range connection via movable ion qubits as in the recent experiment of 4D loop toric code \cite{berthusen2024experiments4dsurfacecode}.   This twisted compatification perspective can also explain the recent discovery that the just-in-time (JIT) decoding scheme which simulates the transversal CCZ or $T$ gates in  3D color/surface code with a (2+1)D spacetime using a 2D setup \cite{Bombin:2018wj, Browneaay4929} is in fact equivalent to the 2D non-Abelian $D_8$ topological order \cite{Davydova:2025ylx}. However, in the JIT scheme \cite{Bombin:2018wj, Browneaay4929}  and equivalently the 2D non-Abelian
code scheme \cite{Davydova:2025ylx}, one only trades space for time without any reduction on the $O(d^3)$ space-time overhead, while our scheme of performing logical CCZ with the 5D self-correcting memory reduces the space-time overhead to $O(d^{5/2})$.   

In future works, our current non-Clifford gate scheme can potentially be further extended to a scheme for a universal logical gate set on self-correcting quantum memories. One can use 4D self-correcting toric codes to implement all  the logical Clifford gates and then potentially perform a single-shot code switching to the 5D non-Abelian codes to perform logical CCZ gates, where the single-shotness is possible due to the self-correcting properties in both types of codes.  We note that such a scheme could outperform the constant-time universal logical gate scheme by Bombin either using the code switching between 3D and 2D color codes \cite{Bombin:2015jk} or the self-correcting scheme using the 6D color code \cite{Bombin:2013cv}, where the space-overhead would be improved from $O(d^3)$ to $O(d^{5/2})$.

From another perspective of  fault-tolerant quantum computation,  a fundamental question is about the space-time overhead for the computation, including the overhead for both the quantum and classical operations in the computing process.  In a typical case of an actively corrected quantum memory, one measures the error syndrome and then sends the syndrome information to the classical computer to decode it. The classical decoder then decides a recovery operation to correct the errors. These processes need non-local classical communications and the classical decoder requires computation time scaled with the system size. Since classical communication and gates are not infinitely faster than the quantum gates, such classical time overhead will make it impossible for the classical decoder to keep up with the advancing of quantum operations and to correct the errors timely when the system is large enough \cite{campbell2017, Brown_2016}.  Self-correcting memory is hence more desirable since it gets rid of the classical software and the associated time overhead and instead passively protects the quantum hardware from the errors.  While such a passive protection might be challenging at the current stage from the engineering perspective, self-correcting memory is also useful even in the context of active error correction since it is always closely associated with an underlying local cellular-automaton decoder \cite{Dennis:2001nw, Brown_2016}.  Such a decoder is composed of local update rules at each time step and does not require non-local classical communication.  It also does not have a decoding time overhead during the computation process except the final readout stage when the classical decoder no longer needs to catch up with the quantum operations.  Although existing single-shot error correction schemes permit  constant quantum time overhead \cite{Bombin:2015hia},  only self-correcting quantum memory can potentially achieve both constant quantum and classical time overhead.    In fact, there exists a scheme for implementing universal logical gate set with constant-depth geometrically non-local circuits acting on a non-Abelian topological code \cite{Zhu:2020_constant_depth, Zhu:2018CodeLong, Lavasani2019universal} where the time overhead is expected to be $O(d/ \log d)$ ($d$ is the code distance) due to the stretching of the support of error clusters by a constant factor.  If such a scheme can be adapted to a non-Abelian self-correcting memory where the stretched error can be shrunk back in constant time, constant time overhead will be achieved for both the quantum and classical operations during the computation stage.

Non-Abelian self-correcting quantum memories correspond to novel non-Abelian topological orders without particles.
Nevertheless, non-Abelian topological orders above $D=4$ spacetime dimension are less understood except for topological orders with particles (see e.g. \cite{Chen:2021xuc}), where the particles arise from the electric charges due to gauging an ordinary symmetry.
On the other hand, topological orders without particles are highly constrained: the lowest-dimensional excitations must obey Abelian fusion rule (see e.g. \cite{Johnson-Freyd:2020ivj,Johnson-Freyd:2021tbq,Cordova:2023bja}). For instance, if the lowest-dimensional nontrivial excitation is a membrane, then it must obey Abelian fusion rule. On the other hand, when there exist nontrivial loop excitations, the membrane excitations do not need to obey Abelian fusion, and they can be non-Abelian.
We will show the following statement regarding TQFTs without particles:
\begin{itemize}
    \item TQFTs without particles can only exist in spacetime dimension $D\geq 5$.

    \item Non-Abelian TQFTs without particles can only exist in spacetime dimension $D\geq 6$.
\end{itemize}
We will derive the statements using two properties:
 remote detectability of TQFTs \cite{PhysRevX.8.021074,Johnson-Freyd:2020ivj} and consistency of fusion rule under shrinking the extended excitations  \cite{PhysRevX.8.021074,Johnson-Freyd:2020ivj,Johnson-Freyd:2021tbq,Cordova:2023bja}.

\subsection{The model: non-Abelian Cubic Theory}

In this work we will construct a family of Non-Abelian self-correcting quantum memories. This is done by gauging the global symmetry of certain class of symmetry-protected topological (SPT) phases with higher-form symmetries in $D\ge 5+1$ spacetime dimensions. This construction is regarded as a generalization of (2+1)D non-Abelian $\mathbb{D}_8$ gauge theory, which is obtained by gauging the SPT phase with $\Z_2^3$ symmetry.

We start with the SPT phase protected by $\mathbb{Z}_2$ $(l-1)$-form symmetry, $(m-1)$-form symmetry and $(n-1)$-form symmetry in $D$ spacetime dimensions, with $l+m+n=D$. The response action for the SPT phase is given by
\begin{equation}
    \pi\int A_l\cup B_m\cup C_{n}~,
\end{equation}
where $A_l,B_m,C_{n}$ are the background gauge fields. In Appendix \ref{sec:cupproductrev} we review the definition of  cup product $\cup$.

After gauging the $\Z_2^3$ symmetry of the above SPT phase, we obtain certain class of non-Abelian topological order.
We will call them generalized $\mathbb{D}_8$ theories or the cubic theories.\footnote{
Despite the naming similarity, the theories are not related to Haah's cubic code.
}
They are described by $\mathbb{Z}_2$ $l$-form, $m$-form and $n$-form gauge theories in $D$ spacetime dimensions, with the topological action
\begin{equation}
    \pi\int a_l\cup b_m\cup c_{n}~,
\end{equation}
where $a_l,b_m,c_{n}$ are the respective dynamical gauge fields.
The theory describes non-Abelian topological order: the fusion rules of the magnetic operators that create magnetic flux for $a_l,b_m,c_{n}$ are non-invertible. For instance, fusing the $(D-l-1)$-dimensional magnetic membrane operator for the gauge field $a_l$ gives
\begin{equation}
\begin{split}
    &M^{(1)}(M_{D-l-1})\times M^{(1)}(M_{D-l-1})= \\
    &\sum_{\gamma_m,\gamma_{n}'}W^{(2)}(\gamma_m)W^{(3)}(\gamma'_{n})~,
    \end{split}
    \label{eq:nonabelianM}
\end{equation}
where $W^{(2)}=(-1)^{\int b_m}$ and $W^{(3)}=(-1)^{\int c_{n}}$ are Wilson surface operators, and the sum is over $\gamma_m\in H_{m}(M_{D-l-1}),\gamma'_{n}\in H_{n}(M_{D-l-1})$. The right hand side is a sum of simple objects, and thus the fusion is non-Abelian. 
When $l=m=n=1, D=3$, the above topological order gives a $(2+1)$D $\mathbb{D}_8$ gauge theory \cite{Yoshida_global_symmetry_2016, Iqbal:2023wvm} with non-Abelian fusion rule of anyons \eqref{eq:nonabelianM}. 
In addition, we also show that the magnetic excitations obey non-Abelian braiding, including Borromean ring type braiding 
of three different magnetic excitations given by a sign--which is a signature of non-Abelian topological order.

For each member in the family, we construct commuting non-Pauli stabilizer lattice models on any triangulated spatial lattice, and in particular on hypercubic lattice.
We call such non-Pauli stabilizer models magic stabilizer codes.

We will show that when $l,m,n$ obey $2\leq l,m,n$, the theory does not have particle excitations. Thus the models provide a family of infinitely many non-Abelian topological order without particles in $D\geq 6$, such as $D=6$ and $l=m=n=2$, where the theory describes Abelian loop excitations and non-Abelian membrane excitations. We will argue that the theories provide thermally stable quantum memory.

\subsection{Review of quantum memory at finite temperature for loop toric code}

Let us review the basic property of self-correcting quantum memory discussed in Ref.~\cite{Dennis:2001nw} for the example of loop-only $\Z_2$ toric code in (4+1)D following a Peierls argument. The Hilbert space has physical qubit on each face of the 4d cubic lattice, and the stabilizer Hamiltonian is
\begin{equation}
    H=H_\text{Gauss}+H_\text{Flux}=-\sum_e \prod_{e\in \partial f} X_f-\sum_c \prod_{f\in\partial c}Z_f~.
\end{equation}
The basic excitations are loop excitations created by membrane operators $\prod X_f$ and $\prod Z_f$.
Take the error given by electric loop excitation of length $\ell$ measured in the lattice spacing, which violates $\ell$ of the Gauss law terms in the Hamiltonian and have energy cost $E(\ell)=2\ell\epsilon_0$ where $\epsilon_0$ is the energy unit of the Hamiltonian. There are $n(\ell)$ such configuration of length $\ell$, and they are suppressed by the Boltzmann factor $e^{-\beta E(\ell)}$.
Whether large errors are suppressed at low temperature depends on the competition of the this entropy effect and the Boltzmann suppression.
The large loop excitation errors are suppressed when
\begin{equation}
    e^{-\beta E(\ell)}n(\ell)\ll 1~.
\end{equation}
where $\beta=1/(k_BT)$ for temperature $T$ and Boltzmann constant $k_B$.
The multiplicity of loops of length $\ell$ is
\begin{equation}
    n(\ell)\sim \text{Polynomial}(\ell)\mu^\ell~,
\end{equation}
where $\mu\sim 6.77$ is the connectivity on 4D hypercubic lattice \cite{madras2012self}, and only self-avoiding loops are counted by resolving the intersection points without loss of generality.
At low temperature, the large loops are exponentially suppressed. The critical temperature below which the errors are suppressed is
\begin{equation}
    T_c\sim (2/\log\mu)\epsilon_0/k_B~.
\end{equation}

A more rigorous argument for the self-correcting property using the Lindbladian evolution under Pauli noise model is discussed in e.g. \cite{Bombin_2013}, and we will use the method there to prove the self-correcting property of our non-Abelian quantum memory.

The work is organized as follows. In Section \ref{sec:obstructionnonabelian} we describe an obstruction to non-Abelian extended excitations in TQFTs.
In Section \ref{sec:lattice} we discuss a family of non-Pauli stabilizer codes with only loops and non-Abelian membrane excitations. 
In Section \ref{sec:D8TQFT} we discuss a family of infinitely many non-Abelian TQFTs in any dimension dubbed a Cubic theory, which effectively describes our lattice model. 
In Section \ref{sec:compactify} we see that the Cubic theory is obtained from the compactification of the toric code in one higher dimension.
In Section \ref{sec:stringmembranes} we describe the fusion rules of the excitations in the (5+1)D lattice model, and describe the braiding involving non-Abelian excitations.
 In Section \ref{sec:CCZ} we construct the fault-tolerant logical CCZ gate of the Cubic theory in (5+1)D.
In Section \ref{sec:self-correctingquantummemory} we argue that our model provides the self-correcting quantum memory at finite temperature. 
In Appendix \ref{sec:cupproductrev} we summarize some properties of cup product on triangulated or hypercubic lattices.
In Appendix \ref{sec:loopTC} we discuss the logical CZ, $S$ and $H$ gates in 4D loop toric code. In Appendix \ref{sec:5D_Clifford} we describe logical CZ, $S$ gates in 5D Cubic theory.

\section{Universal Bounds on Spacetime Dimension for TQFTs Without Particles}
\label{sec:obstructionnonabelian}

Self-correcting quantum memories that are fully topological correspond to TQFTs without particle excitations. Thus we will begin with general discussion of such TQFTs.
In this section we will show the following two statements:
\begin{itemize}
    \item If the TQFTs only have extended excitations, the spacetime dimension has to be at least 5, $D\geq 5$.

    \item If furthermore, some of the extended excitations obey non-Abelian fusion, then the spacetime dimension has to be at least 6, $D\geq 6$.
    
\end{itemize}

In the second statement, we consider excitations that are not condensation descendants \cite{Gaiotto:2019xmp}. This excludes excitations such as $e$-condensed excitation in 4D loop toric code similar to Cheshire string.

The argument uses the following two properties of TQFTs:
\begin{itemize}
    \item Remote detectability \cite{PhysRevX.8.021074,Johnson-Freyd:2020ivj}: an excitation must braid with another excitation.

\item Consistency of fusion under shrinking the excitation on circle  \cite{PhysRevX.8.021074,Johnson-Freyd:2020ivj,Johnson-Freyd:2021tbq,Cordova:2023bja}. We will elaborate more about this in Section \ref{sec:shrinking}.

\end{itemize}

\subsection{Dimension lower bound for particle-free TQFTs}

\begin{figure}[t]
    \centering
    \includegraphics[width=0.75\linewidth]{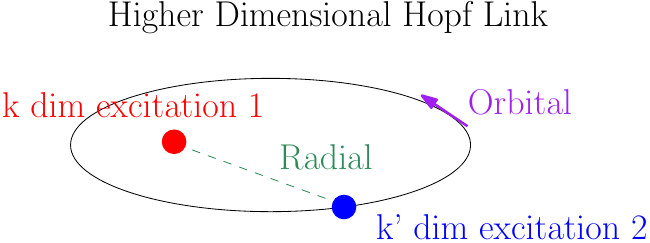}
    \caption{Braiding between $k$ dimensional and $k'$ dimensional excitations in $D$ spacetime dimension requires $k+k'+2=D-1$, where there are 2 dimensions from the radial and orbital directions. The two excitations are extended in extra orthogonal $k,k'$ directions not shown in the figure, respectively, so the two excitations appear as two points that braid with each other on the 2D plane. The blue excitation moves around the red one and braids with each other.}
    \label{fig:braiding}
\end{figure}

For relativistic TQFTs, an excitation of dimension $k$ in $D$ spacetime dimension can braid with another excitation of dimension $k'=D-1-k-2=D-k-3$. The total dimensions for these two excitations together with the radial direction and orbital direction add up to the spatial dimension $(D-1)$, see Fig.~\ref{fig:braiding}.
Thus if the TQFT only has extended excitation,
\begin{equation}
    k\geq 1,\; k'=D-k-3\geq 1\; \Rightarrow \; D-4\geq k\geq 1\; \Rightarrow  \; D\geq 5~.
\end{equation}
In other words, the spacetime dimension $D$ has to be at least $5$. An example is the 4D loop toric code.

\subsection{Dimension lower bound for particle-free non-Abelian TQFTs}
\label{sec:shrinking}

Let us generalize the argument in \cite{Johnson-Freyd:2020ivj,Johnson-Freyd:2021tbq,Cordova:2023bja} that rule out TQFTs with non-Abelian loop excitations in the absence of particles. 
We will use the shrinking method that has been used extensively in the studies of TQFTs above two spatial dimensions, see e.g. \cite{PhysRevX.8.021074,Johnson-Freyd:2020ivj,Johnson-Freyd:2021tbq,Cordova:2023bja}.

\subsubsection{Consistency of fusion under shrinking}

Consider TQFTs without $(n-1)$-dimensional excitations, where we take $n\geq 2$.
Suppose there are $n$-dimensional simple excitations $p_i$ with non-Abelian fusion rules on $S^1\times S^{n-1}$:
\begin{equation}\label{eqn:fusioncontradict}
    p_1\times p_2= \sum_{i=3,4,\cdots}  p_i~,
\end{equation}
where $p_i$ can be the same when the multiplicity is greater than one.

We can shrink the $S^{1}$ to a point, then the $n$-dimensional excitations become $(n-1)$-dimensional excitations on $S^{n-1}$. For simple $n$-dimensional excitations, the reduction on a circle gives an excitation that contains at most one copy of identity. Since there is no nontrivial $(n-1)$-dimensional excitations, this gives only the identity. The fusion rule (\ref{eqn:fusioncontradict}) becomes
\begin{equation}
    1\times 1=\sum_{i=3,4,\cdots} 1~.
\end{equation}
Thus we have a contradiction unless there is only one term on the right hand side. In other words, the fusion of the $n$-dimensional excitations is Abelian.

The only exception for the argument is when $n=1$, i.e. the excitations are particles, which cannot be shrunk further. Indeed, there are TQFTs with only non-Abelian particles.

\subsubsection{Lower bound on spacetime dimension}

Suppose the TQFT has extended non-Abelian excitation of dimension $k$. From the consistency of fusion under shrinking, there must also be Abelian excitation of dimension $\ell=(k-1)$. For the TQFT to satisfy remote detectability, there must also be excitations of dimension $k'=D-k-3$, $\ell'=D-\ell-3=D-k-2$. Demanding the excitations are extended gives
\begin{align}
    &k\geq 1,\quad \ell=k-1\geq 1,\cr 
    &k'=D-k-3\geq 1,\quad \ell'=D-k-2\geq 1\cr 
    &\Rightarrow k\geq 2,\; D-4\geq k\; \Rightarrow\; D\geq 6~.
\end{align}
In other words, the spacetime dimension has to be at least 6.

\section{Cubic Theory in General Spacetime Dimensions}

\subsection{Non-Pauli Stabilizer Hamiltonian}

\label{sec:lattice}

Let us construct a local commuting non-Pauli stabilizer Hamiltonian for the Cubic theory in $D=d+1$ spacetime dimensions. We can begin by constructing a Hamiltonian model for the SPT phase with $\mathbb{Z}_2$ $(l-1)$-form symmetry, $(m-1)$-form symmetry, and $(n-1)$-form symmetry with $l+m+n=D$, and then gauge these symmetries minimally.

\subsubsection{Hamiltonian for SPT phase}

We consider a hypercubic lattice in $d$-dimensional Euclidean space. We introduce a qubit on each $(l-1)$-dimensional hypercube, $(m-1)$-dimensional hypercube, $(n-1)$-dimensional hypercube, acted by Pauli operators $\{\tilde X^{i},\tilde Y^{i},\tilde Z^{i}\}$ with $1\le i\le 3$ respectively.
We use a tilde on the Pauli operators to distinguish them from the Pauli operators in the gauged model we will present later.
Denote the $\Z_2$ variables
\begin{align}
    \lambda^i=(1-\tilde Z^{i})/2~,
    \label{eq:lambda}
\end{align}
which will be later regarded as the $\Z_2$ valued ``matter'' fields transformed under $\Z_2$ global symmetries.
In the basis of Pauli $Z$ operators,
$\lambda^i$ is naturally regarded as $\Z_2$ $p$-cochains $(p=l-1,m-1,n-1)$ for $i=1,2,3$, namely a function from the $p$-dimensional hypercubes to the $\Z_2$ values $\{0,1\}$.

Let us now describe the Hamiltonian for the SPT phase with the topological response action given by
\begin{equation}
    \pi\int A_l\cup B_m\cup C_{n}~,
\end{equation}
with $A_l,B_m,C_{n}$ the background gauge fields. This is a higher-form generalization of the group cohomology SPT phases discussed in \cite{Chen:2011pg}, corresponding to the higher-form analogue of ``type III cocycle''  SPT phase \cite{deWildPropitius:1995cf,Yoshida:2015cia}.
We will follow similar method in \cite{Chen:2011pg} to construct the Hamiltonian. Other examples of lattice models for higher-form SPT phases constructed in this way are discussed in e.g. \cite{Tsui:2019ykk,Chen:2021xks}.

The SPT Hamiltonian is most simply expressed by utilizing the operation of cochains called cup product, denoted by $\cup$. 
For more details, see e.g. \cite{milnor1974characteristic,Benini:2018reh, PhysRevB.101.035101,Chen:2021ppt,Chen:2021xuc}.

\begin{figure}[t]
    \centering
    \includegraphics[width=0.95\textwidth]{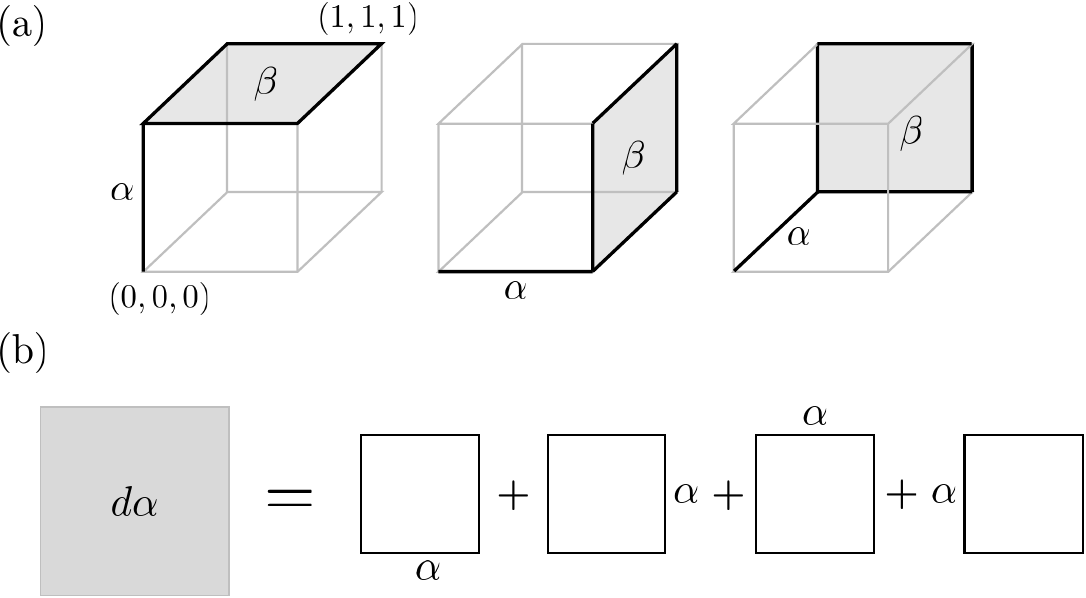}
    \caption{(a): The cup product $\alpha_1\cup \beta_2$ evaluated on a 3d cube. It sums over the possible sequence of a 1d edge and a 2d square sharing a single vertex, starting at $(0,0,0)$ and ending at $(1,1,1)$. (b): Coboundary $d\alpha$ of a $\Z_2$ 1-cochain $\alpha$ on a 2d square is given by summing over $\alpha$ on edges bounding a square.}
    \label{fig:cubecup}
\end{figure}

On hypercubic lattice, the cup product $\alpha_i\cup \beta_j$
of $\mathbb{Z}_2$ $i$-cochain $\alpha_i$ and $j$-cochain $\beta_j$ produces a $(i+j)$-cochain, and its value on $(i+j)$-dimensional hypercube $s_{i+j}$ that span the coordinates $(x^1,x^2,\cdots x^{i+j})\in [0,1]^{i+j}$ is given by
    \begin{equation}
    \begin{split}
        &\alpha_i\cup \beta_j(s_{i+j})= \\
        &\sum_I \alpha_i\left([0,1]^I\right)\beta_j\left((x^I=1,x^{\bar I}=0)+[0,1]^{\bar I}\right)~,
        \label{eq:cup}
        \end{split}
    \end{equation}
    where the summation is over the different sets $I$ of $i$ coordinates out of the $(i+j)$ coordinates $x^1,\cdots x^{i+j}$, $\bar I$ denotes the remaining $j$ coordinates. 
    In each term in the sum, $\alpha_i,\beta_j$ are evaluated on an $i$-dimensional hypercube and an $j$-dimensional hypercube, where the two hypercubes share a single vertex $(x^I=1,x^{\bar I}=0)$:
    \begin{itemize}
        \item $[0,1]^I$ is the $i$-dimensional unit hypercube starting from $(x^I=0,x^{\bar I}=0)$ and ending at $(x^I=1,x^{\bar I}=0)$, i.e. $[0,1]^I=\{0\leq x^\mu\leq 1,x^\nu=0:\mu\in I, \nu\in \bar I\}$.
        \item     $(x^I=1,x^{\bar I}=0)+[0,1]^{\bar I}$ is the $l$-dimensional hypercube in the $\bar I$ directions starting from $(x^I=1,x^{\bar I}=0)$ and ending at $(x^I=1,x^{\bar I}=1)$, i.e. $(x^I=1,x^{\bar I}=0)+[0,1]^{\bar I}=\{x^\mu=1,0\leq x^\nu\leq 1:\mu\in I,\nu\in \bar I\}$.
    \end{itemize}
    For instance, the sum in the cup product \eqref{eq:cup} for $i=1,j=2$ is described in Fig.~\ref{fig:cubecup}. It sums over the possible sequence of two hypercubes sharing a single vertex, starting at $(0,0,\dots 0)$ and ending at $(1,1,\dots 1)$.

    Now we are ready to describe the SPT Hamiltonian. It is obtained by conjugating the trivial SPT Hamiltonian by an entangler. The trivial Hamiltonian is given by
\begin{align}
    H^0=-\sum_{s_{l-1}} \tilde X^{1}_{s_{l-1}}-\sum_{s_{m-1}} \tilde X^{2}_{s_{m-1}}-\sum_{s_{n-1}} \tilde X^{3}_{s_{n-1}}
\end{align}
    where $s_k$ are $k$-dimensional hypercubes. We then consider the entangler
\begin{equation} \label{eq:entangler}
    U= (-1)^{\int \lambda_{l-1}^1\cup d\lambda_{m-1}^2\cup d\lambda_{n-1}^3}~,
\end{equation}
where $d$ is the coboundary of $\Z_2$ cochains; for a given $\Z_2$ $k$-cochain $\alpha$, $d\alpha$ is a $\Z_2$ $(k+1)$-cochain defined as
\begin{align}
    d\alpha(s_{k+1}) = \sum_{s_k\in\partial s_{k+1}} \alpha(s_k).
    \label{eq:coboundary}
\end{align}
See Fig.~\ref{fig:cubecup} (b) for an illustration with $k=1$.
Recall that $\lambda^i$ is regarded as an operator-valued  $\Z_2$ cochains in the basis of Pauli $Z$ operators, and then the cochain $\lambda_{l-1}^1\cup d\lambda_{m-1}^2\cup d\lambda_{n-1}^3$ is a $d$-cochain. The integral in Eq.~\eqref{eq:entangler} is over $d$-chains in the whole space. 
One can re-express the operator-valued cochain as $\lambda_{k}^i $$ = $$ \sum_{\tilde{s}_{k}}\hat{N}^i  \tilde{s}_{k}$ in Eq.~\eqref{eq:entangler}, where $\hat{N}^i=(1-\tilde Z^{i})/2$ is the quantum operator and the classical variable $\tilde{s}_{k}$ is a $k$-cochain that takes value 1 on a single hypercube $s_k$ and zero otherwise. The above entangler in Eq.~\eqref{eq:entangler} can hence be re-written in terms of quantum gates as:
\begin{align}
\nonumber U =& (-1)^{\int \hat{N}^1 \tilde{s}_{l-1}\cup  \hat{N}^2 d\tilde{s}_{m-1} \cup \hat{N}^3 d \tilde{s}_{n-1}}    \\
\nonumber =& \prod_{\tilde{s}_{l-1}, \tilde{s}_{m-1}, \tilde{s}_{n-1}}  \text{CCZ}_{1,2,3}^{\int \tilde{s}_{l-1}\cup  d\tilde{s}_{m-1} \cup d \tilde{s}_{n-1}},
\end{align}
where we have used the gate identity $(-1)^{ \hat{N}^1 \hat{N}^2 \hat{N}^3}= \text{CCZ}_{1,2,3}$ since $\text{CCZ}_{1,2,3}\ket{N^1,N^2,N^3}=(-1)^{N^1 N^2 N^3}\ket{N^1,N^2,N^3}$ ($N^i\in \{0,1\}$ is the eigenvalue of $\hat{N}^i$).  Note the exponent $\int \tilde{s}_{l-1}\cup  d\tilde{s}_{m-1} \cup d \tilde{s}_{n-1} \in \{0,1\}$ is a classical $\ZZ_2$ variable which determines whether the CCZ gate on qubits supported on hypercubes $\tilde{s}_{l-1}$, $\tilde{s}_{m-1}$ and $\tilde{s}_{n-1}$ is applied (value 1) or not (value 0).

\begin{figure}[t]
    \centering
    \includegraphics[width=0.85\textwidth]{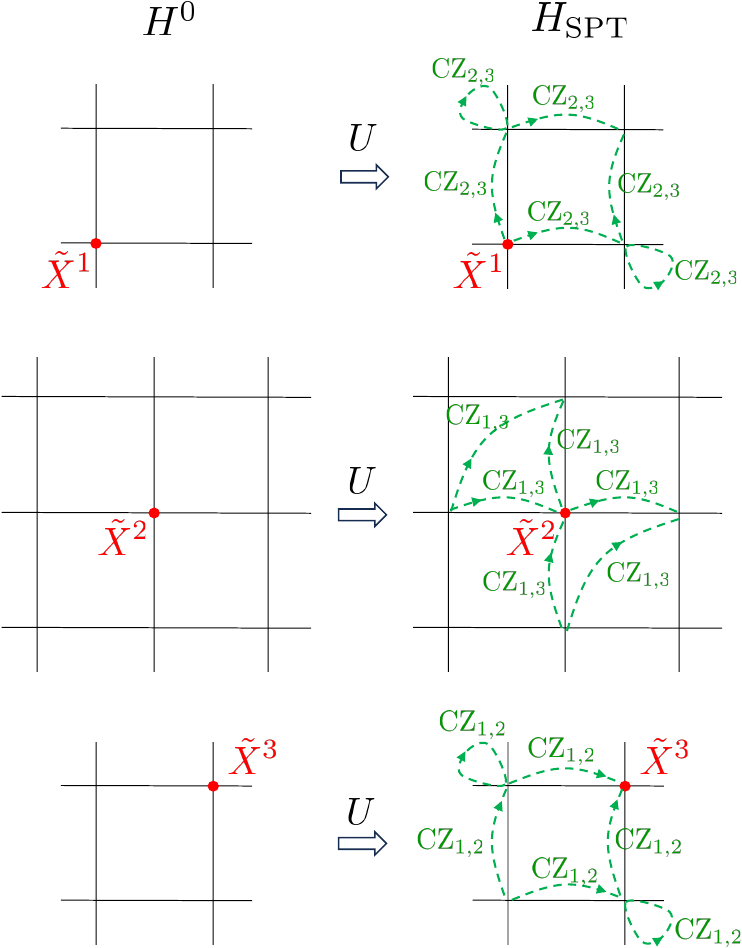}
    \caption{The SPT Hamiltonian is obtained by conjugating the trivial SPT Hamiltonian $H^0$ by the SPT entangler $U$. The figure shows the $\Z_2^3$ SPT phase in (2+1)D, with $l=m=n=1$. The local Hamiltonian term involves the six CZ gates support at a pair of sites at the boundaries of a green dashed curve. The arrow in the dashed curve for $\text{CZ}_{j,k}$ specifies that the outgoing site has the qubit $Z^j$ and the ingoing one has the qubit $Z^k$. This SPT Hamiltonian corresponds to the expression in \eqref{eq:SPT_Hamiltonian} with $l=m=n=1$ in terms of cup product. }
    \label{fig:SPT}
\end{figure}

The SPT Hamiltonian is given by $H_{\text{SPT}}$ obtained by $H_{\text{SPT}} = UH^0 U^\dagger$,
    \begin{align}\label{eq:SPT_Hamiltonian}
    \begin{split}
    H_\text{SPT}= &
    -\sum_{s_{l-1}} \tilde X^{1}_{s_{l-1}}(-1)^{\int \tilde s_{l-1}\cup d\lambda_{m-1}^2\cup d\lambda_{n-1}^3} \\
    &-\sum_{s_{m-1}} \tilde X^{2}_{s_{m-1}}(-1)^{\int  d\lambda_{l-1}^1\cup \tilde s_{m-1}\cup d\lambda_{n-1}^3} \\
    &-\sum_{s_{n-1}} \tilde X^{3}_{s_{n-1}}
    (-1)^{\int d\lambda_{l-1}^1\cup d\lambda_{m-1}^2\cup  \tilde s_{n-1}}  \\
    =& -\sum_{s_{l-1}} \tilde X^{1}_{s_{l-1}}\prod_{\tilde{s}_{m-1}, \tilde{s}_{n-1}} \text{CZ}_{2,3}^{\int \tilde s_{l-1}\cup d\tilde{s}_{m-1}^2\cup d\tilde{s}_{n-1}^3} \\
    &-\sum_{s_{m-1}} \tilde X^{2}_{s_{m-1}} \prod_{\tilde{s}_{l-1},  \tilde{s}_{n-1}} \text{CZ}_{1,3}^{\int  d\tilde{s}_{l-1}^1\cup \tilde s_{m-1}\cup d\tilde{s}_{n-1}^3} \\
    &-\sum_{s_{n-1}} \tilde X^{3}_{s_{n-1}}
   \prod_{\tilde{s}_{l-1}, \tilde{s}_{m-1}}  \text{CZ}_{1,2}^{\int d\tilde{s}_{l-1}^1\cup d\tilde{s}_{m-1}^2\cup  \tilde s_{n-1}},  
   \end{split}
\end{align}
where we have derived the second equality by rewriting the operator-valued cochain as $\lambda_{k}^i $$ = $$ \sum_{\tilde{s}_{k}}\hat{N}^i  \tilde{s}_{k}$ as before, and used the identity $(-1)^{ \hat{N}^i \hat{N}^j}= \text{CZ}_{i,j}$.   The exponent again determines the support of the CZ gates. See Fig.~\ref{fig:SPT} for the illustration of the SPT Hamiltonian with $l=m=n=1$.

This Hamiltonian $H_{\text{SPT}}$ describes a nontrivial SPT phase with $\Z_2^{(l-1)}\times\Z_2^{(m-1)}\times\Z_2^{(n-1)}$ symmetry, where $\Z_2^{(p)}$ denotes $\Z_2$ $p$-form symmetry. Each symmetry is generated by a unitary operator
\begin{align}
    \begin{split}
        \Z_2^{(l-1)}: & \quad \prod_{\tilde s_{l-1}\subset \gamma_{D-l}} \tilde{X}_{s_{l-1}}, \\
        \Z_2^{(m-1)}: & \quad \prod_{\tilde s_{m-1}\subset \gamma_{D-m}} \tilde{X}_{s_{m-1}}, \\
         \Z_2^{(n-1)}: & \quad \prod_{\tilde s_{n-1}\subset \gamma_{D-n}} \tilde{X}_{s_{n-1}}, \\
         \label{eq:globalsym}
    \end{split}
\end{align}
where each $p$-form symmetry operator $(p=l-1,m-1,n-1)$ is supported at a closed $(D-p-1)$ submanifold $\gamma_{D-p-1}$ on a dual lattice of the hypercubic lattice. The product is over the hypercubes $\tilde s_{p}$ that intersects with $\gamma_{D-p-1}$.

\subsubsection{Cubic theory Hamiltonian from Gauging}

We now describe how to gauge the global $\mathbb{Z}_2$ symmetries of our SPT Hamiltonian $H_{\text{SPT}}$. This procedure promotes the global symmetry~\eqref{eq:globalsym} to a local one, by introducing new degrees of freedom—$\mathbb{Z}_2$ gauge fields—and modifying the Hamiltonian by imposing the Gauss law. Our approach follows standard methods developed in Refs.~\cite{Yoshida_global_symmetry_2016,PhysRevB.86.115109,Bhardwaj:2016clt,Shirley:2018vtc}.

\begin{figure*}[t]
    \centering
    \includegraphics[width=0.8\textwidth]{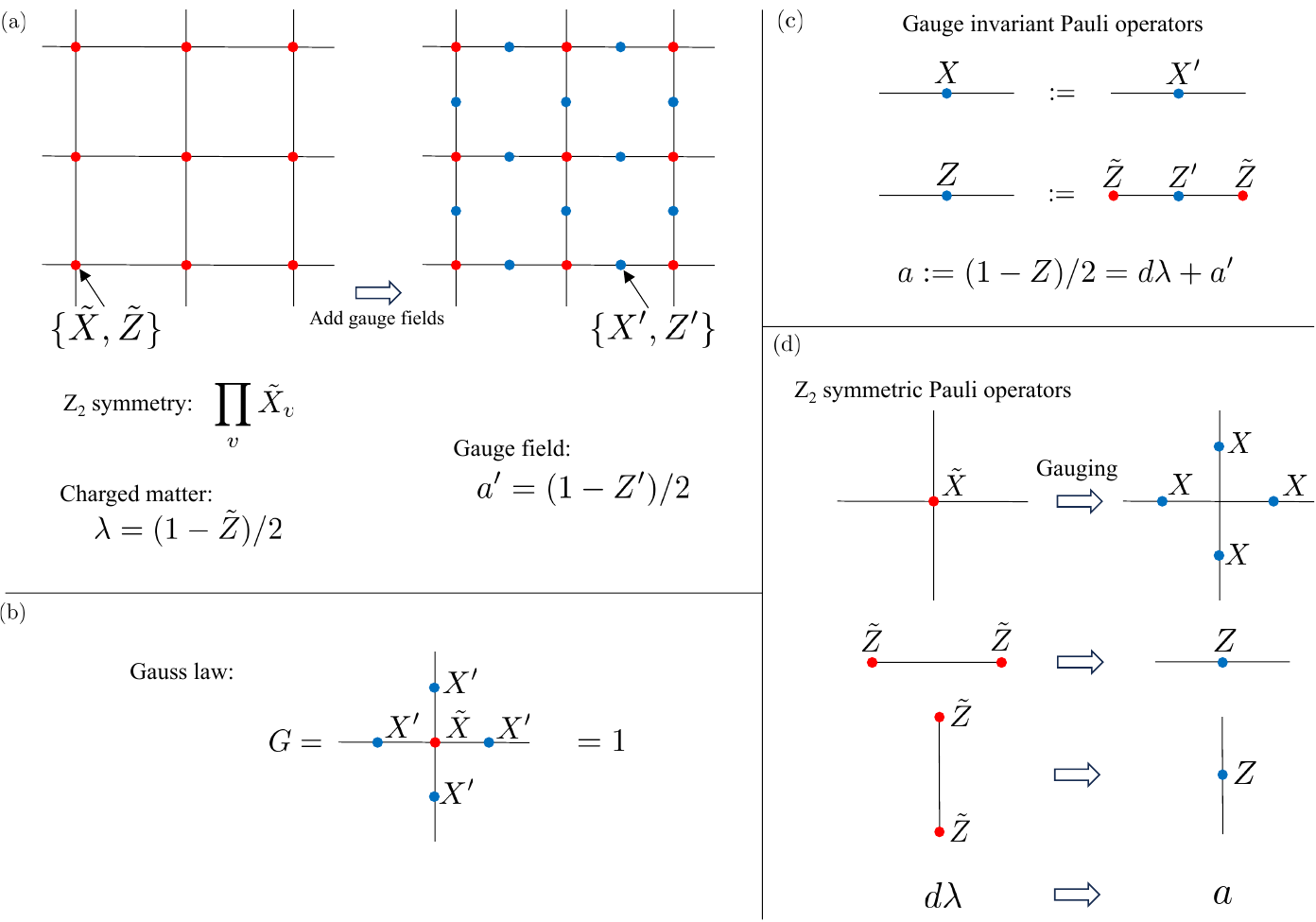}
    \caption{The process of gauging $\Z_2$ 0-form symmetry in (2+1)D (i.e., $l=m=n=1$). (a): We initially have a model with $\Z_2$ 0-form symmetry with charged matter fields $\lambda=(1-\tilde Z)/2$ on each $(l-1)$-hypercubes, which are vertices when $l=1$. The first step is to add the $\Z_2$ gauge fields (blue qubits $\{X',Z'\}$) on $l$-hypercubes $s_l$, which are edges when $l=1$. The $\Z_2$ gauge field is expressed as $a'=(1-Z')/2$. (b): The next step is to impose the Gauss law to define the gauge transformation, and the physical Hilbert space $\mathcal{H}_{\text{phys}}$ is characterized by the gauge invariant states with $G=1$. (c): The gauge invariant operators are the ones commuting with the Gauss law. We define the gauge invariant Pauli operators $\{X,Z\}$, and new $\Z_2$ gauge field $a=(1-Z)/2$, which is related to $a'$ by a gauge transformation.  (d): After minimally coupling the original $\Z_2$ symmetric Pauli Hamiltonian to gauge fields in a gauge invariant manner, the Hamiltonian is expressed in terms of these gauge invariant Pauli operators $\{X,Z\}$. This gauged Hamiltonian can be obtained by the map shown in the figure; a single Pauli $\tilde X$ is mapped to a product of $X$ due to the Gauss law, and the symmetric product of $\tilde Z$ (which is associated with $d\lambda$) is mapped to a single $Z$ (which is associated with the gauge field $a$), due to minimal gauge invariant coupling.}
    \label{fig:gauging}
\end{figure*}

To begin, we introduce qubits representing $\mathbb{Z}_2$ gauge fields. These are placed on the $l$-, $m$-, and $n$-dimensional hypercubes. We denote the Pauli operators acting on these gauge qubits by $X'^i,Y'^i,Z'^i$ with $i=1,2,3$ corresponding to the three types of gauge fields we introduce.
The $\Z_2$ gauge fields correspond to the Pauli $Z'$ eigenvalues of the above qubits:
\begin{align}
    \begin{split}
        a'_l(s_l) &= (1-Z'^1_{s_l})/2, \\
        b'_m(s_m) &= (1-Z'^2_{s_m})/2, \\
        c'_n(s_n) &= (1-Z'^3_{s_n})/2, \\
    \end{split}
\end{align}
which define $\Z_2$ $(p+1)$-cochains ($p=l-1,m-1,n-1$) on a hypercubic lattice. See Fig.~\ref{fig:gauging} (a) for the case $l=m=n=1$.

The idea of gauging is that the matter fields $\lambda_i$ in \eqref{eq:lambda} (transformed under the $\Z_2$ global symmetry \eqref{eq:globalsym} before gauging) get transformed under local $\Z_2$ \textit{gauge} transformations. 
The $\Z_2$ gauge transformations are specified by generic $\Z_2$ cochains $\alpha_{l-1}, \beta_{m-1}, \gamma_{n-1}$; under the gauge transformation, the matter fields are transformed by
\begin{align}
    \lambda^1_{l-1}&\rightarrow \lambda^1_{l-1}-\alpha_{l-1}~,\cr 
    \lambda^2_{m-1}&\rightarrow \lambda^2_{m-1}-\beta_{m-1}~,\cr
    \lambda^3_{n-1}&\rightarrow \lambda^3_{n-1}-\gamma_{n-1}~,
    \label{eq:gaugetrans1}
\end{align}
and the gauge fields are transformed by
\begin{align}
    a'_l &\rightarrow a'_l+d\alpha_{l-1}~,\cr
    b'_m &\rightarrow b'_m+d\beta_{m-1}~,\cr
    c'_{n} &\rightarrow c'_n+d\gamma_{n-1}~,
    \label{eq:gaugetrans2}
\end{align}
where the definition of the coboundary $d$ is given in \eqref{eq:coboundary}.
See Fig.~\ref{fig:cubecup} (b) for an illustration.

The above gauge transformations \eqref{eq:gaugetrans1}, \eqref{eq:gaugetrans2} are generated by the ones with $\alpha_{l-1}= \tilde s_{l-1}, \beta_{m-1} = \tilde s_{m-1}, \gamma_{n-1}=\tilde s_{n-1}$, i.e., the gauge transformation that acts on a single matter field $\lambda^1$ at the hypercube $s_{l-1}$. These gauge transformations are generated by the following Gauss law operators,
\begin{align}
    G^1(s_{l-1}) = &\tilde X^1_{s_{l-1}}\prod_{s_{l-1}\subset \partial s_l} X'^1_{s_l}~,\cr
    G^2(s_{m-1}) =&\tilde X^2_{s_{m-1}}\prod_{s_{m-1}\subset \partial s_m} X'^2_{s_m}~,\cr
    G^3(s_{n-1}) = &\tilde X^3_{s_{n-1}}\prod_{s_{n-1}\subset \partial s_n} X'^3_{s_n}~.
\end{align}
For instance, one can see that $G^1(s_{l-1})$ acts by transforming $\lambda^1_{l-1}\to \lambda^1_{l-1}-\tilde s_{l-1}$, $a'_l\to a'_l + d\tilde s_{l-1}$.

The gauging is performed by imposing the Gauss law constraint $G^1=1, G^2=1, G^3=1$ for each hypercube $s_{l-1}, s_{m-1}, s_{n-1}$. See Fig.~\ref{fig:gauging} (b) for illustrations with $l=m=n=1$.
These Gauss law constraints characterize the gauge invariant ``physical'' Hilbert space $\mathcal{H}_{\text{phys}}$ spanned by the gauge invariant states $G^j\ket{\text{phys}} = \ket{\text{phys}}$. 

We then minimally couple the Hamiltonian $H_{\text{SPT}}$ with the $\Z_2$ gauge fields so that it commutes with the Gauss law operators and acts within the physical Hilbert space $\mathcal{H}_{\text{phys}}$. This is done by 
replacing $d\lambda^1_{l-1}$ in \eqref{eq:SPT_Hamiltonian} by $d\lambda^1_{l-1}+a'_l$, $d\lambda^2_{m-1}$ by $d\lambda^2_{m-1}+b'_m$, $d\lambda^3_{n-1}$ by $d\lambda^3_{n-1}+c'_n$. This defines a gauged Hamiltonian involving the qubits $X'^i,Z'^i,\tilde X^i, \tilde Z^i$.

Let us now define the new Pauli operators on each hypercube $s_l,s_m,s_n$ which commute with the Gauss law operators:
\begin{align}
    \begin{split}
        X^1[s_l] &= X'^1[s_l]~, \\
        Z^1[s_{l}] &= (-1)^{(d\lambda_{l-1} + a'_l)[s_{l}]}~, \\
        X^2[s_m] &= X'^2[s_m]~, \\
        Z^2[s_{m}] &= (-1)^{(d\lambda_{m-1} + b'_m)[s_{m}]}~, \\
        X^3[s_n] &= X'^3[s_n]~, \\
        Z^3[s_{n}] &= (-1)^{(d\lambda_{n-1} + c'_n)[s_{n}]}~. \\
        \label{eq:gaugeinvariant_paulis}
    \end{split}
\end{align}
See Fig.~\ref{fig:gauging} (c) for illustrations of gauge invariant Pauli operators with $l=m=n=1$.

Let us consider the flat $\Z_2$ gauge fields $da'_l=0, db'_m=0, dc'_n=0$. The gauged Hamiltonian for $H_{\text{SPT}}$ is then given in terms of $X^i, Z^i$ by
\begin{align}
    H_\text{Cubic}&=H_\text{Gauss}+H_\text{Flux}~,\cr
    H_\text{Gauss}&=-\sum_{s_{l-1}} \left(\prod_{s_{l-1}\subset \partial s_l} X^1_{s_{l}}\right)\prod_{s_m',s_n'}\text{CZ}_{2,3}^{\int\tilde s_{l-1}\cup \tilde s_m'\cup \tilde s_n'} \cr   
    &\quad -\sum_{s_{m-1}}\left(\prod_{s_{m-1}\subset \partial s_m} X^2_{s_{m}}\right)\prod_{s_l',s_n'}\text{CZ}_{1,3}^{\int \tilde s_l'\cup \tilde s_{m-1}\cup  \tilde s_{n}'}\cr 
    &\quad  -\sum_{s_{n-1}}\left(\prod_{s_{n-1}\subset \partial s_n} X^3_{s_{n}}\right)\prod_{s_l',s_m'}\text{CZ}_{1,2}^{\int \tilde s_l'\cup \tilde s_m'\cup \tilde s_{n-1}}
    ~,
\end{align}
where we rewrite $\tilde X$ term in $H_{\text{SPT}}$ into a product of $X$ using the Gauss law constraints.
This corresponds to the gauged Hamiltonian for $H_{\text{SPT}}$ within the Gauss law constraints $G^i=1$.
Then
\begin{equation}
\begin{split}
    H_\text{Flux} = &-\sum_{s_{l+1}}\left(\prod_{s_{l}\subset \partial s_{l+1}}Z_{s_l}^1\right) 
     -\sum_{s_{m+1}}\left(\prod_{s_{m}\subset \partial s_{m+1}}Z_{s_m}^2\right)\\
 &-    \sum_{s_{n+1}}\left(\prod_{s_{n}\subset \partial s_{n+1}}Z_{s_{n}}^3\right)~,
 \label{eq:Hflux}
\end{split}
\end{equation}
energetically enforces the flat $\Z_2$ gauge fields satisfying  $da'_l=0, db'_m=0, dc'_n=0$ on each hypercube $s_{l+1},s_{m+1}, s_{n+1}$.

As we introduced a new set of gauge invariant Pauli operators in \eqref{eq:gaugeinvariant_paulis}, let us redefine new $\Z_2$ gauge fields as
\begin{align}
    \begin{split}
        a_l[s_l] &= (1-Z^1_{s_l})/2 = (d\lambda_{l-1} + a'_l)[s_l], \\
        b_m[s_m] &= (1-Z^2_{s_m})/2 =(d\lambda_{m-1} + b'_m)[s_{m}], \\
        c_n[s_n] &= (1-Z^3_{s_n})/2=(d\lambda_{n-1} + c'_n)[s_{n}], \\
    \end{split}
\end{align}
$a_l,b_m,c_m$ are related to $a'_l, b'_m, c'_n$ by the gauge transformations with $\alpha_{l-1} = \lambda_{l-1}, \beta_{m-1} = \lambda_{m-1}, \gamma_{n-1} = \lambda_{n-1}$ in \eqref{eq:gaugetrans2} that eliminates the matter fields. Using these new $\Z_2$ gauge fields, the gauged Hamiltonian is expressed as
\begin{align}
    H_\text{Cubic}&=H_\text{Gauss}+H_\text{Flux}~,\cr
    H_\text{Gauss}&=-\sum_{s_{l-1}} \left(\prod_{s_{l-1}\subset \partial s_l} X^1_{s_{l}}\right)(-1)^{\int \tilde s_{l-1}\cup b_m\cup c_{n}} \cr   
    &\quad -\sum_{s_{m-1}}\left(\prod_{s_{m-1}\subset \partial s_m} X^2_{s_{m}}\right)(-1)^{\int a_l\cup \tilde s_{m-1}\cup  c_{n}}\cr 
    &\quad  -\sum_{s_{n-1}}\left(\prod_{s_{n-1}\subset \partial s_n} X^3_{s_{n}}\right)(-1)^{\int a_l\cup b_m\cup \tilde s_{n-1}}~,
\end{align}
and $H_{\text{flux}}$ in \eqref{eq:Hflux} enforces the flatness $da_{l} =0, db_m=0, dc_n=0$.

\begin{figure}[t]
    \centering
    \includegraphics[width=0.8\linewidth]{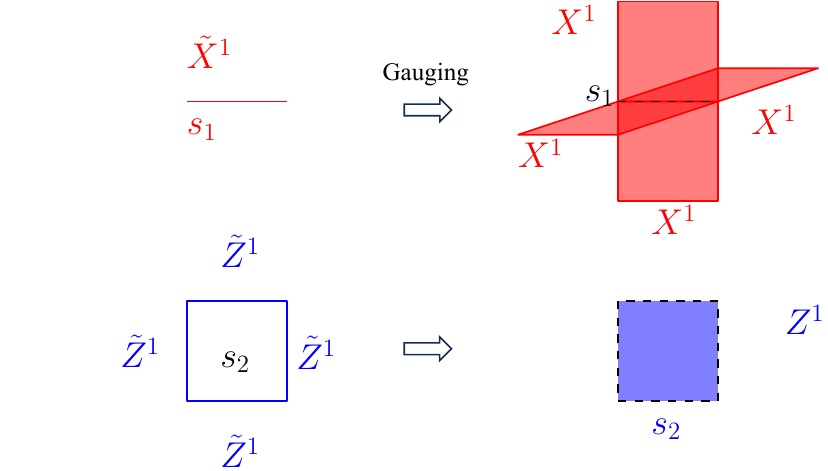}
    \caption{Mapping of operators under the gauging operation in table \ref{tab:gauging} for $l=2$ on 3D cubic lattice. Here $s_1$ and $s_2$ represent the 1-  and 2-dimensional hypercubes, i.e., an edge and a face respectively.   }
    \label{fig:gauging1form}
\end{figure}

In summary, the gauging is regarded as the mapping between Pauli operators formed by $\{\tilde X, \tilde Z\}$ and $\{X,Z\}$ as shown in Table \ref{tab:gauging}. The mapping of Pauli $X$ operators follows from the Gauss law constraints $G^i=1$, and the mapping of Pauli $Z$ operators follows from gauge invariant coupling of operators so that it commutes with the Gauss law. See Fig.~\ref{fig:gauging} (d) for illustrations of mapping of Pauli operators under gauging in the case of $l=m=n=1$. See also Fig.~\ref{fig:gauging1form} for the case with $l=2$ in 3D cubic lattice.

\begin{table}[]
\renewcommand*{\arraystretch}{1.5}
\centering
\begin{tabular}{|c c c|}
\hline
SPT & $\xrightarrow{\text{Gauge }\Z_2^3}$ & Cubic theory \\ \hline
$\tilde X_{s_{l-1}}^{1}$ & $\leftrightarrow$  & $\prod_{s_{l-1}\subset \partial s_l} X^1_{s_{l}}$ \\
$\prod_{s_{l-1}\subset \partial s_{l}} \tilde Z_{s_{l-1}}^{1}$ & $\leftrightarrow$ & $Z_{s_l}^1$ \\ 
$d\lambda^1_{l-1},d\lambda^2_{m-1},d\lambda^3_{n-1}$ & $\leftrightarrow$ & $a_l,b_m,c_n$\\
\hline
\end{tabular}
\caption{The $\Z_2^3$ gauging induces the above mapping between the Pauli operators and cochains. Similar relations also hold for $\{X^2,Z^2\},\{X^3,Z^3\}$. See Fig.~\ref{fig:gauging1form} for an illustration for $l=2$ on 3D cubic lattice.}
\label{tab:gauging}
\end{table}

\begin{figure}[t]
    \centering
    \includegraphics[width=0.85\textwidth]{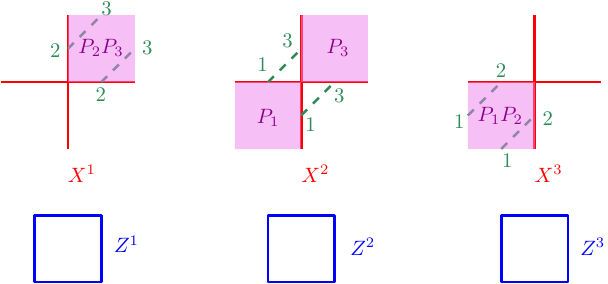}
    \caption{Non-Pauli stabilizer model for $D=3,l=m=n=1$ Cubic theory, which is equivalent to ordinary $\mathbb{D}_8$ gauge theory in (2+1)D. Here $P_i(f)=\frac{1+\prod_{e\in\partial f} Z^i_e}{2}$ is the projector to zero flux on face $f$.
    The dashed line with endpoints $i,j$ is the operator $CZ_{i,j}$ on the qubits $i,j$ on the edges connected by the line.
    The upper figures can multiply with addition flux projectors and still maintain the commuting property.}
    \label{fig:D8stabilizer}
\end{figure}

\subsubsection{Commuting non-Pauli stabilizer lattice model}

We note that the Gauss law terms $H_\text{Gauss}$ only commute among themselves on the zero flux sector. To make them exactly commute with each other, we can modify the Gauss law term by multiplying them with projectors to zero fluxes. 
We note that the flux projectors commute with each term in $H_\text{Gauss}$, and thus the ordering of the projectors do not matter.
This gives commuting Gauss law Hamiltonian:

\tiny{
\begin{align}
\begin{split}
\label{eqn:modifiedGauss}
    &H'_\text{Gauss}=\cr 
    &-\sum_{s_{l-1}} \left(\prod_{s_{l-1}\subset \partial s_l} X^1_{s_{l}}\right)(-1)^{\int \tilde s_{l-1}\cup b_m\cup c_{n}}\cr
    &\quad  \prod_{s'_{m-1}} \left(\frac{1
    +
    (-1)^{\int \tilde s_{l-1}\cup \tilde s_{m-1}'\cup dc_{n}}
    }{2}\right)
    \prod_{s''_{n-1}} \left(\frac{1
    +
    (-1)^{\int \tilde s_{l-1}\cup db_m\cup \tilde s''_{n-1}}}{2}\right)
    \cr   
    &-\sum_{s_{m-1}}\left(\prod_{s_{m-1}\subset \partial s_m} X^2_{s_{m}}\right)(-1)^{\int a_l\cup \tilde s_{m-1}\cup  c_{n}}\cr
    &\quad  \prod_{s'_{l-1}} \left(\frac{1
    +
    (-1)^{\int \tilde s'_{l-1}\cup \tilde s_{m-1}\cup dc_{n}}
    }{2}\right)
    \prod_{s''_{n-1}} \left(\frac{1
    +
    (-1)^{\int da_l \cup \tilde s_{m-1} \cup \tilde s''_{n-1}}}{2}\right)
    \cr       
    & -\sum_{s_{n-1}}\left(\prod_{s_{n-1}\subset \partial s_n} X^3_{s_{n}}\right)(-1)^{\int a_l\cup b_m\cup \tilde s_{n-1}}\cr
    &\quad  \prod_{s'_{l-1}} \left(\frac{1
    +
    (-1)^{\int \tilde s'_{l-1}\cup db_m\cup  \tilde s_{n-1}}
    }{2}\right)
    \prod_{s''_{m-1}} \left(\frac{1
    +
    (-1)^{\int da_l \cup \tilde s''_{m-1} \cup \tilde s_{n-1}}}{2}\right)~.
    \end{split}
\end{align}
}
\normalsize
Here we list several important properties of the Cubic theory Hamiltonian:

\paragraph{Locality of Hamiltonian}
Each term in the Hamiltonian (\ref{eqn:modifiedGauss}) is local. This is because on hypercubic lattice, for a given simplex $s_j$ and integers $k,k'$, $\tilde s_j\cup \tilde s_k\neq 0,\tilde s_{k'}\cup \tilde s_j\neq 0$ for finite number of simplices $s_k,s_{k'}$ independent of the system size.
More explicitly, the number of $s_k$ is given by choosing $k$ positive spatial directions in the complementary $(D-j-1)$ spatial directions, which is $\left(
\begin{array}{c}
     D-j-1\\ k
\end{array}
\right)$. Similarly, the number of $s_{k'}$ is $\left(
\begin{array}{c}
     D-j-1\\ k'
\end{array}
\right)$. For example, if $k=D-j-1$ or $k'=D-j-1$, there is a unique choice of $s_k$ or $s_{k'}$.

\paragraph{Local commuting non-Pauli stabilizer Hamiltonian}
The modified Hamiltonian
\begin{equation}\label{eq:commuting_stabilizer}
    H'_\text{Cubic}=H_\text{Gauss}'+H_\text{Flux}
\end{equation}
is a sum of local commuting terms and has the same ground states as the Hamiltonian $H_\text{Cubic}$.
We remark that the cup product expression allows us to define the model on any lattice that admits a triangulation.
For example, when $D=3,l=m=n=1$, the model is illustrated in Fig.~\ref{fig:D8stabilizer}.
The lattice model defines a non-Pauli stabilizer code describing non-Abelian topological orders, and we will call it magic stabilizer code.

\paragraph{Excitations}

Since the Hamiltonian is a sum of local commuting terms, we can analyze the errors for each term independently, just as in the Calderbank-Shor-Steane (CSS) codes.

Since the Hamiltonian terms commute with each other, the excited states are eigenstates of the Hamiltonian terms. The Gauss law terms can have eigenvalues in $\{-1,0,1\}$, while the flux terms can have eigenvalues in $\{-1,1\}$. There are electric and magnetic excitations, labelled by the eigenvalues of the Hamiltonian terms.
The basic electric excitations are given by violations of the Gauss law term by $1-(-1)=2$ but not the flux term.
The basic magnetic excitations are given by violations of the flux terms by $1-(-1)=2$ and violation of the Gauss law terms by $0-(-1)=1$.

\subsection{Cubic Theory in General Dimension as Non-Abelian TQFT}
\label{sec:D8TQFT}

\subsubsection{Field Theory}

Here we describe the gauge theory for the Cubic theory in $D$ spacetime dimensions.
Consider $\mathbb{Z}_2$ $l$-form, $m$-form and $n$-form gauge fields $a_l,b_m,c_n$ satisfying $l+m+n=D$, with the action
\begin{equation}\label{eqn:topaction}
    \pi\int a_l\cup b_m\cup c_{n}~.
\end{equation}
We can write the theory by embedding the gauge fields into $U(1)$ gauge fields:
\small
\begin{equation}\label{eqn:U(1)norm}
    \frac{1}{\pi^2}\int a_l b_m c_{n}+\frac{2}{2\pi}\int (a_l d\tilde a_{D-l-1} + b_m d\tilde b_{D-m-1} +  c_n d\tilde c_{D-n-1})~.
\end{equation}
\normalsize

Examples of the theory are discussed in various literature:
\begin{itemize}
    \item When $l=1,m=1$, the gauge fields $(a_1,b_1,\tilde c_1)$ describe the gauge field of $\mathbb{D}_8$ one-form gauge theory: the equation of motion for $c_{D-2}$ sets $d\tilde c_{1}=\frac{1}{\pi}a_1 b_1$, which described the extension of $\mathbb{Z}_2\times\mathbb{Z}_2$ by $\mathbb{Z}_2$ with the 2-cocycle given by $a_1 b_1$.
    When $D=3$, the equivalence with the $\mathbb{D}_8$ gauge theory is discussed in e.g. \cite{DEWILDPROPITIUS1997188,Coste:2000tq,Hsin:2019fhf}.

    \item When $l=2,m=2,$ and $D=5$, this is gauging the $\mathbb{Z}_2$ 0-form symmetry of the (4+1)D 2-form $\Z_2^2$ gauge theory
    \begin{equation}
    \frac{2}{2\pi}\int (a_2 d\tilde a_{2} + b_2 d\tilde b_{2})
\end{equation}    
    generated by the toric code Walker Wang domain wall with the generator given by $\exp(\frac{i}{\pi}\int a_2b_2)$.

\end{itemize}

\subsubsection{Hilbert space of TQFT}

The equation of motions are (here we normalize the gauge fields to have holonomy $0,\pi$ as in (\ref{eqn:U(1)norm}))
\begin{align}
\begin{split}
    &da_l=0,\quad db_m=0,\quad dc_{n}=0\cr
    &d\tilde a_{D-l-1}+b_mc_{n}/\pi=0, \\
    &d\tilde b_{D-m-1}+a_lc_{n}/\pi=0, \\
    &d\tilde c_{D-n-1}+a_lb_m/\pi=0~.
    \end{split}
    \label{eq:EOM}
\end{align}
Thus the Hilbert space of the TQFT on a spatial manifold can be described by holonomies of $a_l,b_m,c_{n}$ subject to the constraints (below we use the normalization that the holonomies are $0,1$ mod 2)
\begin{equation}
    a_l\cup b_m=0,\quad a_l\cup c_{n}=0,\quad b_m\cup c_{n}=0~.
\end{equation}

We note that the constraints on the $\Z_2$ gauge fields described above reproduces those on the ground state of the Cubic theory Hamiltonian derived in Section \ref{sec:groundstate_cubic}.

\subsubsection{Operators}

The theory has the following topological operators: (here we normalize the gauge fields to have holonomy $0,\pi$ as in (\ref{eqn:U(1)norm}))
\begin{itemize}
    \item The theory has invertible operators generated by the electric Wilson operators
\begin{equation}
    W^{(1)}=e^{i\int a_l},\ W^{(2)}=e^{i\int b_m}, \ W^{(3)}=e^{i\int c_n}~.
\end{equation}
The electric Wilson operators obey $\mathbb{Z}_2^3$ fusion rule.

\item The magnetic operators are 
\begin{align}
    &e^{i\int_{\Sigma_{D-l-1}} \tilde a_{D-l-1}+i\int_{{\cal V}_{D-l}} b_mc_{n}/\pi}\cr 
    &e^{i\int_{\Sigma_{D-m-1}} \tilde b_{D-m-1}+i\int_{{\cal V}_{D-m}} a_{l}c_n/\pi}\cr 
    &e^{i\int_{\Sigma_{D-n-1}} \tilde c_{D-n-1}+i\int_{{\cal V}_{D-n}} a_lb_m/\pi}~,
\end{align}
where $\Sigma_k=\partial {\cal V}_{k+1}$. 
Since in the Hilbert space $a_l\cup b_m=0$, $a_l\cup c_{n}=0$, $b_m\cup c_{n}=0$, these operators do not depend on the choice of ${\cal V}_i$ bounded by $\Sigma_{i-1}$ for $i=D-l,D-m,D-n$, respectively \cite{Barkeshli:2022edm}.
To obtain magnetic operators without ${\cal V}_{k+1}$, we can put choose a local polarization \cite{Witten:1996hc,Witten:1998wy,Witten:2009at,Freed:2012bs}, i.e. consider ${\cal V}_{k+1}$ with two boundaries, one with magnetic operators, the other a topological boundary condition. The topological boundary condition can be described by a normal subgroup $H$ of the bulk gauge group such that the topological action on ${\cal V}_{k+1}$ becomes trivial, i.e. well-defined boundary without gauge anomalies, and a topological action on the boundary for the $H$ subgroup.

For instance, we can obtain a magnetic operator for $a_l$ on $\Sigma_{D-l-1}$ by choosing the other topological boundary to be $b_m|=c_{n}|=0$, which gives a magnetic operator decorated with the condensation of the Wilson operators $W^{(2)},W^{(3)}$ on $\Sigma_{D-l-1}$ that imposes the projection to $b_m|=0,c_{n}|=0$ \cite{Cordova:2024defects}. We will call such magnetic operator $M^{(1)}$, and similarly define $M^{(2)},M^{(3)}$ using the Dirichlet boundary conditions of $(a_l,c_{n}),(a_l,b_m)$, respectively.

From the condensate of Wilson operators, one can derive
the non-Abelian fusion rules (see also \cite{Barkeshli:2022edm})
\begin{align}
    &M^{(I)}\times W^{(J)}=M^{(I)}~, I\neq J\cr 
    &M^{(I)}\times M^{(I)}=\sum W^{(J)}(\gamma)W^{(K)}(\gamma')~,
\end{align}
for distinct $I,J,K$. In particular, the fusion rule in the second equation corresponds to that of the magnetic operators in the lattice model \eqref{magneticfusion_lattice}.

\item Gauged symmetry-protected topological (SPT) operators \cite{Yoshida_gate_SPT_2015, Yoshida_global_symmetry_2016, Yoshida2017387, Barkeshli:2022edm, barkeshli2023codimension} $H^*(B^l\mathbb{Z}_2\times B^m\mathbb{Z}_2\times B^{n}\mathbb{Z}_2,U(1))$. 
The closed gauged SPT operators $\int a_l b_m,\int a_l c_{n},\int b_mc_{n}$ are trivial.
There are following nontrivial gauged SPT operators:
\begin{equation}
\begin{split}
 V^{12} &=   e^{i\int a_ldb_m/(2\pi)}, \\
 V^{23} &=e^{i\int b_mdc_{n}/(2\pi)}, \\
V^{13} &=e^{i\int a_l dc_{n}/(2\pi)}~.
 \end{split}
\end{equation}

\item ``Mixed'' operators that involve the electric and the magnetic operators:
\begin{align}
    &e^{i\int_{\partial M_{D-m+l}} a_l \tilde b_{D-m-1}/\pi+i\int_{M_{D-m+l}} a_l^2 c_{n}/\pi^2}\cr 
    &e^{i\int_{\partial M_{D-l+m}} b_m \tilde a_{D-l-1}/\pi+i\int_{M_{D-m+l}} b_m^2 c_{n}/\pi^2}\cr 
    &e^{i\int_{\partial M_{D-n+m}} b_m \tilde c_{D-n-1}/\pi+i\int_{M_{D-n+m}} b_m^2 a_{l}/\pi^2}\cr 
    &e^{i\int_{\partial M_{D-m+n}} c_n \tilde b_{D-m-1}/\pi+i\int_{M_{D-m+n}} c_n^2 a_{l}/\pi^2}\cr 
    &e^{i\int_{\partial M_{D-l+n}} c_n \tilde a_{D-l-1}/\pi+i\int_{M_{D-l+n}} c_n^2 b_{m}/\pi^2}\cr 
    &e^{i\int_{\partial M_{D-n+l}} a_l \tilde c_{D-n-1}/\pi+i\int_{M_{D-n+l}} a_l^2 b_{m}/\pi^2}\cr 
\end{align}
As in the magnetic operators, these operators do not depend on the bulk. To define operators without using the bulk, we can choose a local polarization by putting the operator on one boundary and a topological boundary condition on the other boundary. For instance, in the first operator we can impose the Dirichlet boundary conditions for $a_l,c_{n}$, and similar for the other operators. Let us call the resulting operators on the boundary ${\cal D}^{12},{\cal D}^{21},{\cal D}^{23},{\cal D}^{32},{\cal D}^{31},{\cal D}^{13}$. Due to the condensate from the topological boundary conditions, the operators ${\cal D}^{ij}$ are non-invertible.

The operators act on the Wilson and magnetic operators.
For instance, when the Wilson operator $W^{(2)}$ intersects ${\cal D}^{12}$, it creates holonomy for $\tilde b_{D-m-1}$, and results in additional Wilson operator $W^{(1)}$. On the other hand, when the magnetic operator $M^{(1)}$ intersects ${\cal D}^{12}$, the projectors for $W^{(1)}$ annihilate the magnetic operator, and similarly for $M^{(3)}$. In other words, the magnetic operators $M^{(1)}, M^{(3)}$ cannot intersect with or terminate at $\mathcal{D}^{12}$:
\begin{equation}
\begin{split}
    {\cal D}^{12}: \quad & W^{(1)}\rightarrow W^{(1)},\quad   W^{(2)}\rightarrow W^{(1)}W^{(2)},\\
    & M^{(1)}\rightarrow 0,\quad M^{(3)}\rightarrow 0~.
    \end{split}
\end{equation}

\end{itemize}

\subsubsection{Cubic theories without particles}

Let us consider the class of cubic theories that do not have particles. In other words, all gauge fields $a,b,c,\tilde a,\tilde b,\tilde c$ have degree greater or equal 2. This requires
\begin{equation}
\begin{split}
    & l\geq 2,\quad m \geq 2,\quad n\geq 2, \\
    & D-l-1\geq 2,\quad D-m-1\geq 2,\quad D-n-1\geq 2~.
    \end{split}
\end{equation}
These conditions simplify to
\begin{equation}\label{eqn:noparticles}
    l\geq 2,\quad m \geq 2,\quad n\geq 2~.
\end{equation}
This inequality gives $l+m+n = D\geq 6$.
The condition on spacetime dimension $D$ is consistent with the no-go theorem for non-Abelian TQFTs without particles. The inequality is saturated with $D=6,l=m=n=2$. For every spacetime dimension $\ge$ 6, there is a non-Abelian TQFT without particles given by the Cubic theory.

As a consistency check, we note that in theories without particles, since the electric excitations obey Abelian fusion rule and magnetic excitations obey non-Abelian fusion rules, the dimensions of the magnetic excitations must be strictly greater than the minimal dimension of the electric excitations due to the constraint discussed in Section \ref{sec:obstructionnonabelian} (see also \cite{Johnson-Freyd:2021tbq,Cordova:2023bja}):
\begin{align}\label{eqn:noparticle2}
    &D-l-1> \text{min}(l,m,n)\cr
    &D-m-1> \text{min}(l,m,n)\cr 
    &D-n-1> \text{min}(l,m,n)~.
\end{align}
In such case, we can use electric excitations to form condensate whose dimension is one less than that of the magnetic excitations. Since there are condensates of all dimensions down to the dimension of the electric excitation itself which obeys Abelian fusion rules, the condition in Section \ref{sec:obstructionnonabelian} follows from (\ref{eqn:noparticle2}).
On the other hand, we can reproduce the conditions (\ref{eqn:noparticle2}) from the inequalities in (\ref{eqn:noparticles}) for the cubic theories without particles.
 The inequalities of (\ref{eqn:noparticle2}) are related by permutation of $l,m,n$, so without loss of generality we can consider the first one of them; suppose the minimum on the right hand side is $l$ (including the case when the minimum is degenerate), then $(D-l-1)> l$ follows from $m\ge n$ and $n\geq 2$ in (\ref{eqn:noparticles}). 
    
Thus the theories without particles have the properties that the 
lowest dimensional excitations obey Abelian fusion rule, in agreement with constraint on excitations with non-Abelian fusion rules (see e.g. Section \ref{sec:obstructionnonabelian}).

\subsection{Cubic Theory from Compactifying the Generalized Color Code}
\label{sec:compactify}

The Cubic theory in $D$ spacetime dimensions can be obtained from twisted compactification of a generalized color code theory in
$(D+1)$ spacetime dimensions \cite{Bombin:2013cv, Kubica:2015mta}, where the generalized color code theory is equivalent to decoupled $\mathbb{Z}_2$ $l$-form gauge theory, $\mathbb{Z}_2$ $m$-form gauge theory and $\mathbb{Z}_2$ $n$-form gauge theory. 

\begin{figure}[th!]
    \centering
    \includegraphics[width=1\textwidth]{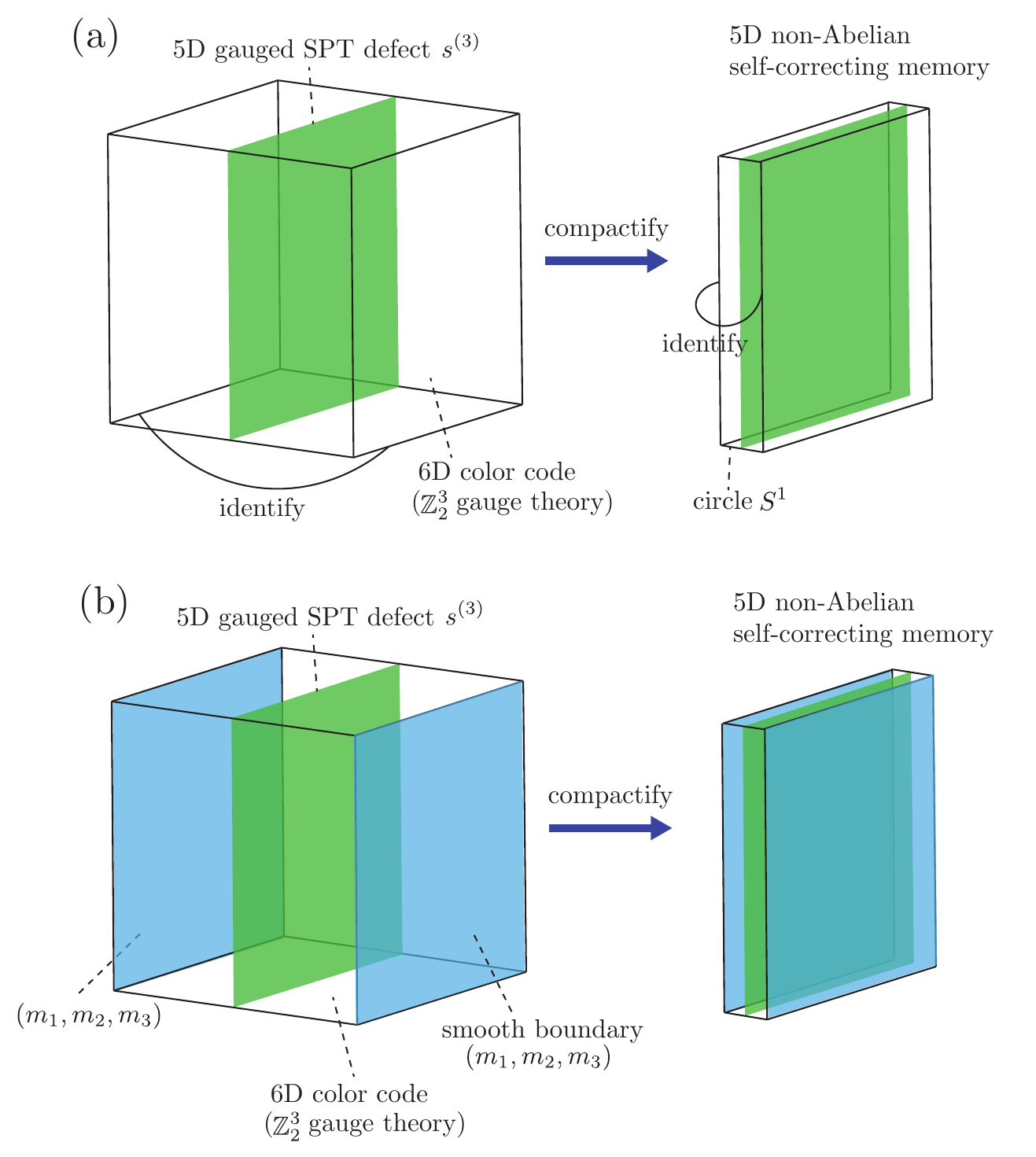}
    \caption{Non-Abelian self-correcting memory from (a) compactification of color code on a circle with gauged SPT defect inserted, and (b) reduction on an interval with $m$-condensed boundaries (i.e. Neumann boundary conditions for the higher-form $\mathbb{Z}_2$ gauge fields) on two ends and the gauged SPT defect inserted in the middle.}
    \label{fig:compactification}
\end{figure}

The $(D+1)$-dimensional theory can be described by three copies of $\mathbb{Z}_2$ toric code with qubits on spatial $l$-hypercubes, $m$-hypercubes and $n$-hypercubes, respectively. 
It is equivalent to a single copy of generalized color code up to a constant-depth local quantum circuit (see e.g. \cite{Kubica:2015mta, Bombin:2006sj, Bombin:2013cv}).  For example, the $D=6, l=m=n=2$ case corresponds to the color code in 6d space, which is a self-correcting memory described in Ref.~\cite{Bombin:2013cv}. This gives rise to our non-Abelian self-correcting memory in five spatial dimensions after the twisted compactification process.

The $(D+1)$ dimensional theory has a $\mathbb{Z}_2$ 0-form symmetry generated by gauged SPT defect (denoted by $s^{(3)}$) \cite{Yoshida_gate_SPT_2015, Yoshida_global_symmetry_2016, Yoshida2017387, Barkeshli:2022edm, barkeshli2023codimension} given by domain wall decorated with the topological action (\ref{eqn:topaction}). If the space is $S^l\times S^m\times S^{n}$, the $\mathbb{Z}_2$ 0-form symmetry acts as a constant-depth circuit composed of the product of CCZ gates on the triplet of qubits in three copies of toric codes respectively, which is dual to the transversal $T$ gate in the generalized color code \cite{Kubica:2015mta}.  The corresponding symmetry operator in the TQFT can be expressed as:
\begin{equation}\label{eq:CCZ defect}
    \mathcal{D}_{s^{(3)}}(M_{D}) = \exp\left( \pi i  \int_{M_{D+1}} a_l \cup b_m \cup c_{n}\right),
\end{equation}
where for the $D=3,l=m=n=1$ case the correspondence to the CCZ gates in three copies of 3D toric codes has been discussed in Ref.~\cite{Yoshida_global_symmetry_2016, barkeshli2023codimension} and the correspondence to the transversal T gates in the 3D color code can be found in Ref.~\cite{zhu2023non}.

We note that when a magnetic defect of one gauge field pierces the domain wall $s^{(3)}$, it forms a junction with the domain wall that has the non-Abelian magnetic defect in the Cubic theory, whose fusion produces the Wilson operators for the other two $\mathbb{Z}_2$ gauge fields. This gives an operator-value associator similar to the junction discussed in section 2.3 of \cite{Barkeshli:2023bta}.

The twisted compactification of the $(D+1)$-dimensional theory to $D$ spacetime dimensions is given by turning on holonomy of this $\mathbb{Z}_2$ 0-form symmetry along the internal circle direction, i.e. placing the domain wall at a point on the circle, and we keep only the zero modes of the gauge fields $a_l,b_m,c_{n}$ that do not wind around the circle.\footnote{
The winding modes will give additional $(l-1)$-form, $(m-1)$-form and $(n-1)$-form gauge fields $\int a_l,\int b_m,\int c_{n}$.
}
The holonomy introduce the action
\begin{equation}
    \pi\int a_l\cup b_m\cup c_{n} \cup (d\theta/2\pi)~,
\end{equation}
where $\theta\sim \theta+2\pi$ is the coordinate of the circle, and this corresponds to the background gauge field $d\theta/2\pi$ for the $\mathbb{Z}_2$ 0-form symmetry. Integrating over the circle direction gives the topological action (\ref{eqn:topaction}). 
Alternatively, we can reduce the $(D+1)$-dimensional theory on an interval with Neumann boundary condition for the gauge fields and the $\mathbb{Z}_2$ domain wall in the middle. The Neumann boundary conditions for the gauge fields correspond to the $m$-condensed or smooth boundaries. See Fig.~\ref{fig:compactification} for an illustration.

We remark that similar twisted dimensional reduction for twisted gauge theory in $D=3$ is discussed in Ref.~\cite{Hsin:2019fhf}.

\section{Cubic Theory in (5+1)D: Loops and Membranes}
\label{sec:stringmembranes}

In the following, we will focus on the Cubic theory with $l=m=n=2$, $D=6$. This is a non-Abelian topological order in (5+1)D with string and membrane excitations. The strings obey Abelian fusion rule \cite{Johnson-Freyd:2021tbq,Cordova:2023bja}, while the membranes are non-Abelian.

\subsection{Commuting non-Pauli stabilizer lattice Hamiltonian model}

Let us recall the commuting non-Pauli stabilizer model for the Cubic theory in (5+1)D.
On each face we introduce three types of qubits for $a_2,b_2,c_2$, acted by Pauli operators $X^i,Y^i,Z^i$. Denote the $\Z_2$ gauge fields $a_2=(1-Z^1)/2,b_2=(1-Z^2)/2,c_2=(1-Z^3)/2$.
The commuting Hamiltonian $H'_{\text{Cubic}}$ is given by
\footnotesize
\begin{align}
\begin{split}
    &H'_{\text{Cubic}}= \cr
    &-\sum_{e} \left(\prod_{\partial f\supset e} X^1_f\right) (-1)^{\int \tilde e\cup b_2\cup c_{2}} \cr
    & \quad\prod_{e'} 
    \left(\frac{1+(-1)^{\int \tilde e\cup \tilde e'\cup dc_2}}{2}\right)
    \prod_{e''} \left(\frac{1+(-1)^{\int \tilde e\cup db_2\cup \tilde e''}}{2}\right)\cr 
    & -\sum_{e} \left(\prod_{\partial f\supset e} X^2_f\right) (-1)^{\int a_2\cup \tilde e\cup c_{2}} \\
    & \quad \prod_{e'} 
    \left(\frac{1+(-1)^{\int \tilde e'\cup \tilde e\cup  dc_2}}{2}\right)
    \prod_{e''} \left(\frac{1+(-1)^{\int da_2\cup \tilde e\cup  \tilde e''}}{2}\right)\cr 
& 
-\sum_{e} \left(\prod_{\partial f\supset e} X^3_f\right) (-1)^{\int a_2\cup b_2\cup \tilde e} \\
& \quad \prod_{e'} 
    \left(\frac{1+(-1)^{\int \tilde e'\cup db_2\cup \tilde e'}}{2}\right)
    \prod_{e''} \left(\frac{1+(-1)^{\int da_2\cup \tilde e''\cup \tilde e}}{2}\right)\cr 
    &-\sum_{c}\left(\prod_{f\subset \partial c}Z_{f}^1\right) -\sum_{c}\left(\prod_{f\subset \partial c}Z_{f}^2\right)-\sum_{c}\left(\prod_{f\subset \partial c}Z_{f}^3\right)~.
    \end{split}
    \label{eq:cubic5d}
\end{align}
\normalsize
where $e,f,c$ denote edges, faces and cubes.
For instance, on the 5d space with coordinate $(x,y,z,u,v)$, the first term without the projector on the edge $(x,y,z,u,v)=(0,0,0,0,0)$ to $(1,0,0,0,0)$ in the $x$ direction is given by product of $X_f^1$ on the 16 faces that span respectively the $\hat x\times (\pm\hat y),\hat x\times (\pm \hat z),\hat x\times (\pm \hat u),\hat x\times (\pm\hat v)$ directions, where $\hat x$ denotes the unit vector in the $x$ direction.
The $(-1)^{\int \tilde e\cup b_2\cup c_2}$ is given by $CZ_{f,f'}$ of the physical qubits that are acted by the Pauli operators $X^2_f,X^3_{f'}$ on the following pair of faces:
\begin{itemize}
    \item $f=(x=1,\text{Square}_{y=0,z=0},u=v=0)$ and $f'=(x=1,y=1,z=1,\text{Square}_{u=0,v=0})$, where $\text{Square}_{y=0,z=0}$ is the unit area square on the $y,z$-plane with four vertices $(y,z)=(0,0),(0,1),(1,0),(1,1)$. We note that the face $f$ meet the edge $(0\leq x\leq 1,y=z=u=v=0)$ at a single point $(x=1,y=z=u=v=0)$, the face $f'$ does not meet this edge, and the faces $f,f'$ meet at a single point $(x=1,y=1,z=1,u=v=0)$.

    \item The other pairs are permutations of $y,z,u,v$, with total $\left(\begin{array}{c}
       4  \\ 2 
    \end{array}
    \right)=6$ pairs of $(f,f')$ in total.
\end{itemize}
The projectors are product of the zero flux projectors $(1+\prod_{f''\in\partial c} Z^2_{f''})/2,(1+\prod_{f'''\in \partial c'} Z^3_{f'''})/2$ over the neighboring cubes $c,c'$.

\subsection{Ground states}
\label{sec:groundstate_cubic}
Let us describe the ground state Hilbert space of the Cubic theory Hamiltonian \eqref{eq:cubic5d}. Due to the terms in $H_{\mathrm{Flux}}$, the $\Z_2$ gauge fields is flat within the ground state,
\begin{align}
    da_2=0, db_2=0, dc_2=0.
\end{align}

In addition, by multiplying the terms of $H_{\text{Gauss}}$ over the edges crossing a closed 4d membrane in the dual lattice, we get another constraints on the $\Z_2$ gauge fields within the ground state,
\begin{align}
    \int_{\Sigma_4} a_2\cup b_2=0, \int_{\Sigma_4} a_2\cup c_2=0, 
    \int_{\Sigma_4} b_2\cup c_2=0, 
    \label{eqn:holonomycondition_lattice}
\end{align}
on any closed 4d closed membrane $\Sigma_4$ embedded in the space. Each ground state is labeled by the configuration of the $\Z_2$ gauge field 
satisfying the above two constraints, up to gauge equivalence.

\subsection{Extended operators on the lattice}

\subsubsection{Wilson surface operators}
On the lattice, there are Wilson membrane operators given by product $\prod Z^i_f$ over each small face on the closed 2d surface embedded in the space. On non-contractible 2-cycles they give rise to nontrivial logical operators.

\subsubsection{Magnetic volume operators}
There are also magnetic volume operators described by product of $X^i$ together with projectors for each $i$. The projector is product $(1+Z^j)(1+Z^k)$ on non-contractible 2-cycles in the volume for distinct $j,k\neq i$ to preserve the condition (\ref{eqn:holonomycondition_lattice}), i.e. project out the states that will violate the condition after applying $X^i$.
For instance, the magnetic operators for the gauge field $c_2$ on 3-cycle $M_3$ are 
\begin{equation}\label{eqn:magneticoperator}
    M^{(3)}(M_3)=P \left[X^3(M_3^\vee)\prod_{f,f'\subset M_3}\left(1+Z^1_f\right)\left(1+Z^2_{f'}\right)\right] P~
\end{equation}
for distinct $j,k\neq i$, where $M_3^\vee$ is the 3-cycle on the dual lattice obtained by half translating the original 3-cycle $M_3$ by a vector $(1/2,\dots,1/2)$, and $X^3(M_3^\vee)$ is the product of $X^3$ over faces of the original lattice crossing $M_3^\vee$. The term $\left(1+Z^1_f\right)\left(1+Z^2_{f'}\right)$ are projectors for $Z^1, Z^2$ support at $M_3$. The other magnetic operators $M^{(1)},M^{(2)}$ are expressed in a similar manner up to slight microscopic modification about the position of the projectors.

The projectors enforce the condition (\ref{eqn:holonomycondition_lattice}) evaluated on $M_3^\vee\times M_2,M_3^\vee\times M_2'$ when the holonomy of the $i$th gauge field is changed on $M_3^\vee$ ($i=a,b,c$).
The operator $P$ denotes the projector onto the ground state subspace, i.e.~the product of projectors for all local stabilizers along $M_3$. The magnetic operators commute with the Hamiltonian due to the projector, and they act on the ground states by changing the eigenvalues of large Wilson operators.

As a consequence of the projector, the magnetic operators are non-invertible, but obey the fusion rule
\begin{equation}
M^{(i)}(M_3)\times M^{(i)}(M_3)=\sum_{M_2,M_2'\in H_2(M_3)} W^{(j)}(M_2)W^{(k)}(M_2') ~,
\label{magneticfusion_lattice}
\end{equation}
for distinct $j,k\neq i$.
Thus the magnetic volume operators describe non-Abelian membrane excitations.

We remark that another way to see the non-Abelian magnetic operator is constructing the non-Abelian Cubic theory from $\mathbb{Z}_2^{(2)}\times\mathbb{Z}_2^{(2)}$ 2-form gauge theory of $a_2,b_2$, i.e. two copies of two-form toric code in (5+1)D, by gauging the $\mathbb{Z}_2$ SPT one-form symmetry \cite{Barkeshli:2022edm,Barkeshli:2023bta} generated by $(-1)^{\int a_2\cup b_2}$, with two-form $\mathbb{Z}_2$ gauge field $c_2$. Such gauging operation couples the $a_2,b_2$ theories. The twist defect, i.e. the magnetic defect of the gauge field $c_2$, lives on the boundary of such SPT and thus carries projective representation \cite{Barkeshli:2014cna,Teo_2015,Tarantino_2016}, which implies the non-Abelian fusion rule (\ref{magneticfusion_lattice}).

\subsection{Code distance}

We remark that on a hypercubic lattice with linear size $L$ in each direction, the minimal logical operator is given by Wilson membrane operators of size $L^2$. The errors created by smaller size operators correspond to excited states. Thus the code distance is $d=O(L^2)=O(N^{\frac{2}{5}})$, where $N=O(L^5)$ represents the total number of qubits.

\subsection{Cubic Theory in (5+1)D as Non-Abelian TQFT}
\label{sec:5dcubic}

Here we present the TQFT descriptions of the Cubic theory and its extended operators in (5+1)D.

\subsubsection{TQFT Hilbert space}

The equation of motion implies
\begin{equation}\label{eqn:holonomycondition}
    a_2\cup b_2=0,\quad a_2\cup c_2=0,\quad b_2\cup c_2=0~. 
\end{equation}
The Hilbert space is described by different holonomies of the two-form $\mathbb{Z}_2$ gauge fields $a_2,b_2,c_2$ subject to the above constraint.

For instance, suppose the space is $S^2\times S^2\times S^1$. Denote the holonomies on the two $S^2$s by the three-component vectors $\vec{n}_i=(n^a_i,n^b_i,n^c_i)$ with $n^a_i=0,1$ mod 2 is the holonomy of $a_2$ on the $i$th $S^2$, the constraint is
\begin{equation}
    \vec{n}_1\times \vec{n}_2=0\text{ mod }2~.
\end{equation}
There are 22 solutions, thus the Hilbert space of the TQFT on $S^2\times S^2\times S^1$ has dimension 22.

\subsubsection{Operators}

There are Wilson surface operators
\begin{equation}
    W^{(1)}=e^{i\int a_2},\quad 
    W^{(2)}=
    e ^{i\int b_2},\quad 
    W^{(3)}=
    e^{i\int c_2}~.
\end{equation}
The integral is over the support of the Wilson surface, given by 2-cycles $M_2$. The Wilson surface operators obey $\mathbb{Z}_2^3$ fusion rule. 
Under union of surfaces, the Wilson operators satisfy $W^{(i)}(M_2)\times W^{(i)}(M_2')=W^{(i)}(M_2+M_2')$ for any two 2-cycles $M_2,M_2'$.
On the lattice, they are described by product of $Z^i$ for each $i$. The Wilson surface operators correspond to string excitations, and they obey $\mathbb{Z}_2^3$ fusion rules.

There are magnetic volume operators
\begin{equation}
\begin{split}
M^{(1)} &=    e^{i\int \tilde a_3+i\int b_2c_2/\pi}, \\
M^{(2)} &=    e^{i\int \tilde b_3+i\int a_2c_2/\pi}, \\
M^{(3)} &=    e^{i\int \tilde c_3+i\int a_2b_2/\pi}~.
\end{split}
\end{equation}
The integral is over the support of the volume operators, given by 3-cycles $M_3$.

\subsubsection{Non-Abelian braiding of magnetic operators}

\paragraph{Two-membrane braiding gives 0.}

The magnetic operators obey non-Abelian braiding following the method of \cite{Barkeshli:2022edm}.
The braiding can be derived from the commutator of the non-invertible operators (\ref{eqn:magneticoperator}).
Consider the correlation function of two different magnetic operators
\begin{equation}
    \langle M^{(1)}(V_3) M^{(2)}(V_3')\rangle~.
\end{equation}
The equation of motion for $\tilde a_3,\tilde b_3$ sets $da_2=-\pi\delta(V_3)^\perp$, 
 $db_2=-\pi\delta(V_3')^\perp$, where $\delta(V_3)^\perp$ is a delta function 3-form that restricts integrals to $V_3$. 
 For $V_3=\partial V_4,V_3'=\partial V_4'$ this reduces to $a_2=-\pi \delta(V_4)^\perp,b_2=-\pi \delta(V_4')^\perp$.
We are left with 
\begin{equation}
    e^{i\int c_2\delta(V_4)^\perp \delta(V_4')^\perp}~.
\end{equation}
The equation of motion for $c_2$ sets $d\tilde c_3=\pi\delta(V_4)^\perp \delta(V_4')^\perp$. For non-exact $\delta(V_4)^\perp \delta(V_4')^\perp$ the equation does not have a solution, and thus the correlation function is zero.
Another way to see the correlation function is zero is by performing path integral over $\mathbb{Z}_2$ gauge field $c_2$, whose possible holonomy is $0,\pi$ mod $2\pi$: for the configuration where $\int c_2\delta(V_4)^\perp \delta(V_4')^\perp=0,\pi$, the correlation function is
\begin{equation}
    e^{0 i}+e^{\pi i}=0~.
\end{equation}
Such vanishing correlation function indicates the magnetic operators are non-Abelian. This generalizes the vanishing Hopf braiding of two non-Abelian particles in (2+1)D $\mathbb{D}_8$ gauge theory (see e.g. \cite{Coste:2000tq}).
This is similar to the braiding of non-Abelian Ising anyon.
The above computation indicates that there are two braiding channels.

In the lattice model on periodic space of $T^5$ topology, consider the commutator of the magnetic operators $M^{(1)}(T^3_{1,2,3}),M^{(2)}(T^3_{1,4,5})$ supported on two 3-tori with subscripts labelling the coordinates, where
\small
\begin{align}
\begin{split}
    M^{(1)}(T^3_{1,2,3})
    &=P \left[X^1(T^{3\vee}_{1,2,3})\prod_{f,f'\subset T^3_{1,2,3}}\left(1+Z^2_f\right)\left(1+Z^3_{f'}\right)\right] P \cr
    &= P X^1(T^{3\vee}_{1,2,3})\prod_{f,f'\subset T^3_{1,2,3}}\left(1+Z^2_f\right)\left(1+Z^3_{f'}\right)  \\
    &\quad \times \left(\frac{1+Z^2(T^2_{2,3})}{2}\right) P \cr
    M^{(2)}(T^3_{1,4,5}) &= P \left[X^1(T^{3\vee}_{1,4,5})\prod_{f,f'\subset T^3_{1,4,5}}\left(1+Z^2_f\right)\left(1+Z^3_f\right)\right] P~,
    \end{split}
    \end{align}
    \normalsize
where $Z^2(T^2_{2,3})$ is the product of $Z^2_f$ over the 2-torus $T^2_{2,3}$.
From the above expressions we find
\begin{align}
\begin{split}
    & M^{(1)}(T^3_{1,2,3})M^{(2)}(T^3_{1,4,5}) \\
        = & M^{(1)}(T^3_{1,2,3})M^{(2)}(T^3_{1,4,5}) \cdot \frac{1-Z^2(T^2_{2,3})}{2}~
        \end{split}
\end{align}
due to the anti-commutation relation between $Z^2(T^2_{2,3})$ and $X^1(T^{3\vee}_{1,4,5})$. Meanwhile we have
\begin{align}
\begin{split}
    & M^{(1)}(T^3_{1,2,3})M^{(2)}(T^3_{1,4,5}) \\
        =& \frac{1+Z^2(T^2_{2,3})}{2} \cdot M^{(1)}(T^3_{1,2,3})M^{(2)}(T^3_{1,4,5}) ~.
        \end{split}
\end{align}
The braiding can then be obtained as
\begin{equation}
    M^{(1)}(T^3_{1,2,3})M^{(2)}(T^3_{1,4,5})M^{(1)}(T^3_{1,2,3})M^{(2)}(T^3_{1,4,5})=0~,
\end{equation}
which follows from $\left(1-Z^2(T^2_{2,3})\right)\left(1+Z^2(T^2_{2,3})\right)=0$,
i.e. $M^{(1)}(T^3_{1,2,3})M^{(2)}(T^3_{1,4,5})$ and $M^{(2)}(T^3_{1,4,5})M^{(1)}(T^3_{1,2,3})$ have orthogonal projectors.

\begin{figure*}
    \centering
    \includegraphics[width=0.7\textwidth]{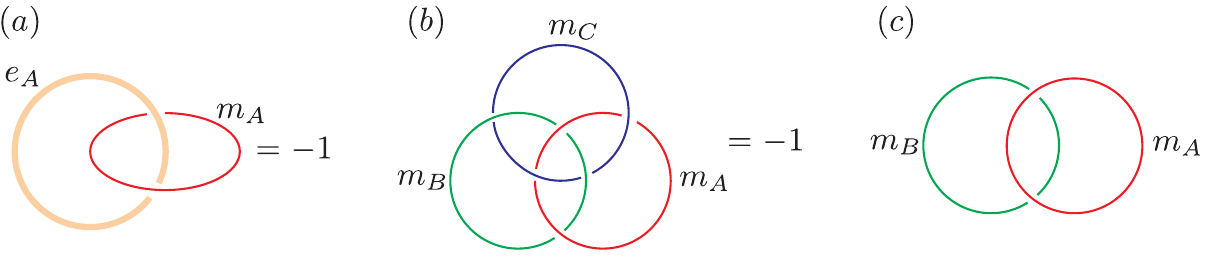}
    \caption{ Illustration of the braiding process in (2+1)D space-time. (a).~Mutual braiding phase ($e^{i \pi}=-1$) of the $e_A$ and $m_A$ particles in the copy $A$ of the (2+1)D $\mathbb{D}_8$ topological order.  (b).~The Borromean ring braiding phase of the $m_A$, $m_B$, and $m_C$ particles in the (2+1)D $\mathbb{D}_8$ topological order.  This braiding process can also be generalized to the (5+1)D where the $m$-excitation becomes 2D membrane.  (c). In the absence of the worldline of $m_C$, there is no linking between the wordlines of $m_A$ and $m_B$ which leads to a trivial phase between them.}
    \label{fig:boromin_ring}
\end{figure*}

\paragraph{``Borromean rings'' braiding of membranes.}

If we include additional magnetic operator $M^{(3)}(V_3'')$, i.e. correlation function of $M^{(1)}(V_3),M^{(2)}(V_3'),M^{(3)}(V_3'')$, the new magnetic operator sources the holonomy of $c_2=-\pi \delta(V_4'')$ with $\partial V_4''=V_3''$, thus the three magnetic operators have the correlation function 
\begin{equation}
    (-1)^{\int \delta(V_4)^\perp\delta(V_4')^\perp \delta(V_4'')^\perp}~.
\end{equation}
The correlation gives $(-1)$ for the three volume operators arranged in analogue of Borromean rings. The discussion is similar to the $D=3,m=n=1$ case in \cite{Putrov:2016qdo}.
See Fig.~\ref{fig:boromin_ring} for illustrations in the (2+1)D model. 
Therefore the non-Abelian membrane excitations, just like the non-Abelian excitations in the $\mathbb{D}_8$ topological orders, enjoy nontrivial Borromean ring type braiding given by a sign, which does not have any pairwise braiding of the membrane excitations-- this implies that the process cannot be realized in Abelian topological orders.

\section{Non-Clifford Logical CCZ Gate from Higher Cup Product}
\label{sec:CCZ}
In this section we will construct fault-tolerant non-Clifford CCZ logical gate for the Cubic theory using constant depth circuit.

\subsection{Logical CCZ gate}

Let us again focus on the spacetime dimensions $D=6$ with $l=m=n=2$, i.e., a 5d non-Abelian self-correcting code. We will see that the Cubic theory in this setup has a non-Clifford logical gate generated by a local finite depth circuit. While we described the Cubic theory Hamiltonian on a hypercubic lattice, here we consider the Hamiltonian based on a generic triangulation (equipped with a branching structure). The qubits $X^i,Z^i$ with $i=1,2,3$ are located on each 2-simplex of the triangulation.
The Cubic theory Hamiltonian can be defined in the completely same manner as the hypercubic lattice:
\begin{align}
    H_\text{Cubic}&=H_\text{Gauss}+H_\text{Flux}~,\cr
    H_\text{Gauss}&=-\sum_{s_{l-1}} \left(\prod_{s_{l-1}\subset \partial s_l} X^1_{s_{l}}\right)(-1)^{\int \tilde s_{l-1}\cup b\cup c} \cr   
    &\quad -\sum_{s_{m-1}}\left(\prod_{s_{m-1}\subset \partial s_m} X^2_{s_{m}}\right)(-1)^{\int a\cup \tilde s_{m-1}\cup  c}\cr 
    &\quad  -\sum_{s_{n-1}}\left(\prod_{s_{n-1}\subset \partial s_n} X^3_{s_{n}}\right)(-1)^{\int a\cup b\cup \tilde s_{n-1}}~,
\end{align}

\begin{equation}
\begin{split}
    H_\text{Flux} = &-\sum_{s_{l+1}}\left(\prod_{s_{l}\subset \partial s_{l+1}}Z_{s_l}^1\right) 
     -\sum_{s_{m+1}}\left(\prod_{s_{m}\subset \partial s_{m+1}}Z_{s_m}^2\right)\\
 &-    \sum_{s_{n+1}}\left(\prod_{s_{n}\subset \partial s_{n+1}}Z_{s_{n}}^3\right)~,
\end{split}
\end{equation}
where the operations $\cup, \partial$ follow the standard definition on a triangulation.
For instance, each term of $H_{\text{Gauss}}$ is expressed as a product of $X$ and CZ operators on generic triangulated manifolds as
\begin{align}
    \begin{split}
        & -\sum_{s_{1}} \left(\prod_{s_{1}\subset \partial s_2} X^1_{s_{1}}\right)(-1)^{ \tilde s_{1}\cup b\cup c} \\
        = &  -\sum_{s_{1}} \left(\prod_{s_{1}\subset \partial s_2} X^1_{s_{1}}\right)\prod_{\substack{ \Delta_5=\langle 012345\rangle \\ s_1 = \langle 01\rangle}} \text{CZ}^{2,3}_{\langle 123\rangle \langle 345\rangle}~,
    \end{split}
\end{align}
\begin{align}
    \begin{split}
        & -\sum_{s_{1}} \left(\prod_{s_{1}\subset \partial s_2} X^1_{s_{1}}\right)(-1)^{\int a\cup \tilde s_{1}\cup c} \\
        = &  -\sum_{s_{1}} \left(\prod_{s_{1}\subset \partial s_2} X^1_{s_{1}}\right)\prod_{\substack{ \Delta_5=\langle 012345\rangle \\ s_1 = \langle 23\rangle}} \text{CZ}^{1,3}_{\langle 012\rangle \langle 345\rangle}~,
    \end{split}
\end{align}
\begin{align}
    \begin{split}
        & -\sum_{s_{1}} \left(\prod_{s_{1}\subset \partial s_2} X^1_{s_{1}}\right)(-1)^{\int a\cup b\cup \tilde s_{1}} \\
        = &  -\sum_{s_{1}} \left(\prod_{s_{1}\subset \partial s_2} X^1_{s_{1}}\right)\prod_{\substack{ \Delta_5=\langle 012345\rangle \\ s_1 = \langle 45\rangle}} \text{CZ}^{1,2}_{\langle 012\rangle \langle 234\rangle}~,
    \end{split}
\end{align}
where the product is over 5-simplices $\Delta_5=\langle 012345\rangle$ that contain $s_1$ with a desired label of simplices $s_1=\langle jk\rangle$.

Then, the logical gate of the Cubic theory is generated by a unitary
\begin{align}
    U = \left(\bigotimes_f (\text{SWAP})^{1,2}_f \right) \times  (-1)^{\int (a\cup_1 b)\cup c }
\end{align}
where $a\cup_1 b$ is higher cup product defined as
\begin{align}
    (a\cup_1 b)(0123) = a(013)b(123) + a(023)b(012)~,
\end{align}
on each 3-simplex (0123). See also Appendix \ref{sec:highercup} for its definition.

The operator $(-1)^{\int (a\cup_1 b)\cup c }$ is the product of CCZ gates, evaluating the cup product $(a\cup_1 b)\cup c$ on each 5-simplex in the space
\begin{align}
    (-1)^{\int (a\cup_1 b)\cup c } = \prod_{\Delta_5=\langle012345\rangle }  \text{CCZ}^{1,2,3}_{\langle013\rangle\langle123\rangle\langle 345\rangle}
\text{CCZ}^{1,2,3}_{\langle023\rangle\langle012\rangle\langle 345\rangle}
\end{align}
This unitary is followed by the transversal SWAP gate swapping the qubits as $X^1\leftrightarrow X^2, Z^1\leftrightarrow Z^2$. 

This operator $U$ becomes an emergent symmetry of the Hamiltonian $H_{\text{Cubic}}$ within the zero flux sector $da=db=dc=0$. Once we write down the non-Pauli commuting stabilizer Hamiltonian $H'_{\text{Cubic}}$ by projecting onto the zero flux sector as we done in \eqref{eq:commuting_stabilizer}, then $U$ becomes an exact symmetry of $H'_{\text{Cubic}}$.

To see how $U$ becomes the logical gate, let us conjugate the Cubic theory Hamiltonian by $U$. $H_{\text{Flux}}$ obviously commutes with $U$, so let us focus on the Hamiltonian $H_{\text{Gauss}}$. By using the properties of cup product, within the zero flux sector $da=db=dc=0$ we have
\begin{align}
    \begin{split}
       &(-1)^{\int (a\cup_1b)\cup c}\left(\prod_{s_{l-1}\subset \partial s_l} X^1_{s_{l}}\right)(-1)^{\int (a\cup_1b)\cup c} \\
       &= \left(\prod_{s_{l-1}\subset \partial s_l} X^1_{s_{l}}\right) (-1)^{\int \tilde s_{l-1}\cup b\cup c + b\cup \tilde s_{l-1} \cup c }~,
    \end{split}
\end{align}
where we used the Pauli $X$ term shifts the gauge field by $a\to a + d\tilde s_{l-1}$, and that
\begin{align}
    d\tilde s\cup_1 b = d(\tilde s\cup_1 b) + \tilde s \cup b + b\cup\tilde s~.
\end{align}
Based on a similar logic, we have
\begin{align}
    \begin{split}
       &(-1)^{\int (a\cup_1b)\cup c}\left(\prod_{s_{m-1}\subset \partial s_m} X^2_{s_{m}}\right)(-1)^{\int (a\cup_1b)\cup c} \\
       &= \left(\prod_{s_{m-1}\subset \partial s_m} X^2_{s_{m}}\right) (-1)^{\int  a\cup \tilde s_{m-1}\cup c + \tilde s_{m-1}\cup a \cup c }~,
    \end{split}
\end{align}
\begin{align}
    \begin{split}
       &(-1)^{\int (a\cup_1b)\cup c}\left(\prod_{s_{n-1}\subset \partial s_n} X^3_{s_{n}}\right)(-1)^{\int (a\cup_1b)\cup c} \\
       &= \left(\prod_{s_{n-1}\subset \partial s_n} X^3_{s_{n}}\right) (-1)^{\int  a\cup b\cup \tilde s_{n-1} + b\cup a \cup \tilde s_{n-1} }~.
    \end{split}
\end{align}
By using the above commutation relations, we get
\begin{align}
    \begin{split}
        & (-1)^{\int (a\cup_1b)\cup c} H_{\text{Gauss}} (-1)^{\int (a\cup_1b)\cup c} \\
        &= -\sum_{s_{l-1}} \left(\prod_{s_{l-1}\subset \partial s_l} X^1_{s_{l}}\right)(-1)^{\int b\cup\tilde s_{l-1}\cup c} \cr   
    &\quad -\sum_{s_{m-1}}\left(\prod_{s_{m-1}\subset \partial s_m} X^2_{s_{m}}\right)(-1)^{\int \tilde s_{m-1}\cup a \cup  c}\cr 
    &\quad  -\sum_{s_{n-1}}\left(\prod_{s_{n-1}\subset \partial s_n} X^3_{s_{n}}\right)(-1)^{\int b\cup a\cup \tilde s_{n-1}}~.
\end{split}
\end{align}
Then, this conjugated Hamiltonian is related to the original one $H_{\text{Gauss}}$ by exchanging the operators as $X^1\leftrightarrow X^2$, $a\leftrightarrow b$ on each 2-simplex. This is realized by the transversal SWAP gate. This shows that $U H_{\text{Cubic}} U^\dagger =H_{\text{Cubic}}$ within zero flux sector, showing that this generates the logical gate of the Cubic theory.

The unitary $U$ generates a non-Clifford logical gate with a specific choice of 5d spatial manifold. For example, take the space to be the Wu 5-manifold $\text{Wu}=SU(3)/SO(3)$. The manifold does not have a spin structure or spin$^c$ structure, thus the Stiefel-Whitney classes $w_2,w_3$ are nontrivial. It is also known that $w_2\cup w_3$ is nontrivial in $\Z_2$ cohomology. Moreover, $H^4(\text{Wu},\mathbb{Z}_2)=1$ and thus $w_2^2$ is trivial in $\mathbb{Z}_2$ cohomology.

Thus the code space is equivalent to three logical qubits, which corresponds to the configurations of gauge fields
\begin{equation}
    a=n_aw_2,\quad b=n_bw_2,\quad c=n_c w_2~,
\end{equation}
where $n_a,n_b,n_c=0,1$ label the logical qubits, and $a\cup b,b\cup c,a\cup c$ are trivial. 

Then, the operator $(-1)^{\int (a\cup_1 b)\cup c}$ evaluates as a logical CCZ gate
\begin{align}
    &(-1)^{\int (a\cup_1 b)\cup c}=(-1)^{n_an_bn_c\int w_3\cup w_2}\cr 
    &\quad =(-1)^{n_an_bn_c}=\overline{\text{CCZ}}_{1,2,3}~,
\end{align}
where we used $w_2\cup_1 w_2 = \text{Sq}^1w_2 = w_3$ on closed oriented manifolds, with $\text{Sq}^1$ the Steenrod operation.
The transversal SWAP gate acts by swapping the first two logical qubits, therefore the logical action of $U$ becomes a logical non-Clifford gate
\begin{align}
    U = \overline{\text{SWAP}}_{1,2} \overline{\text{CCZ}}_{1,2,3}~.
    \label{eq:logicalCCZ}
    \end{align}

The fact that the SWAP symmetry is decorated with the CCZ gate originates from the topological action $\pi\int a\cup b\cup c$. In the absence of the topological action, the automorphism of the gauge group that exchanges $a,b$ is simply given by the SWAP gate. However, the topological action is not exactly invariant under the automorphism: 
\begin{align}
    b\cup a \cup c = a\cup b\cup c + d((a\cup_1 b)\cup c)~.
\end{align}
See Appendix \ref{sec:highercup} for the definition of $\cup_1$ product, and also Appendix A of \cite{Benini:2018reh} for more properties.
Thus the automorphism is not simply the SWAP gate, but SWAP gate decorated with $(-1)^{\int (a\cup_1b)\cup c}$, which gives the logical CCZ gate in the symmetry operator \eqref{eq:logicalCCZ}. In a companion paper we systematically generalize such construction to non-Clifford gates in other quantum codes~\cite{kobayashi2025cliffordhierarchystabilizercodes, hsin2025automorphismgaugetheorieshigher}.

\subsection{Distance scaling beyond $n^{1/3}$ distance barrier}

In this section, we construct a family of non-Abelian self-correcting memory based on the Wu 5-manifold $M^5$. 

Since the Wu manifold is a smooth manifold, it admits a triangulation according to Whitehead and Cairns. We start with a minimum triangulation $\Lambda$ of the Wu manifold containing a finite number ($l$) of 5-simplices (or more generally any triangulation with finite number of 5-simplices).   We then fine-grain the coarse triangulation $\Lambda$ to a fine triangulation $\tilde{\Lambda}$ by subdividing each coarse 5-simplex $\Delta_j$ in $\Lambda$ uniformly into many fine 5-simplices in the following way:  we randomly insert $n \in \mathbb{N}$ new vertices in each coarse 5-simplex $\Delta_j$ such that the density of these $n$ vertices are the same anywhere in $\Delta_j$.  We then build a Delaunay triangulation using these $n$ vertices and the original vertices of the coarse 5-simplex $\Delta_j$.  It is known that for the given set of vertices there is only a unique  Delaunay triangulation.  Also, the Delaunay triangulation guarantees that the degree of the vertices is  bounded so that each $q$-simplex is only connected to a bounded number of $(q \pm 1)$-simplex, and one hence ends up with a quantum LDPC code.  Since the vertices have bounded degree, we end up with $O(n)$ $q$-simplices ($0 \le q \le 5$) in each coarse simplex $\Delta_j$. The qubits in this code  are placed on the 2-simplex, and we hance have in total $l \cdot O(n)$ qubits in the corresponding quantum code.  Note that when increasing $n$, we obtain a family of codes with $O(n)$ qubits (since $l$ is a constant).

The two types of logical operators (Wilson surface and magnetic volume  operators) are supported on non-trivial 2-cycles $\alpha_2 \in H_2(M^5; \ZZ_2)$ and 3-cycles $\alpha_3 \in H_3(M^5; \ZZ_2)$ respectively. We define the combinatorial $\ZZ_2$-$i$-systole as the size of the $i$-cycles with minimum support, i.e.,
\begin{equation}
sys_i(M^5; \ZZ_2) = \{\min(|\alpha_i|) : \alpha_i \in H_i(M^5; \ZZ_2) \},
\end{equation}
where $|\alpha_i|$ counts the number of $i$-simplices in the $i$-cycle $\alpha_i$.
The distance of the code is hence 
\begin{equation}
d=\min(sys_2(M^5; \ZZ_2), sys_3(M^5; \ZZ_2)). 
\end{equation}

Since the new vertices in each coarse 5-simplex $\Delta_i$ are uniformly distributed and the degrees are bounded, the fine  triangulation $\tilde{\Lambda}$ is locally Euclidean-like, i.e., with effectively near-zero curvature in the bulk of the coarse 5-simplex $\Delta_i$.   Now for a given $n$,  the 2-cycle with minimum support $\tilde{\alpha}_2$ intersect with each coarse 5-simplex $\Delta_j$ in $w_j$ fine 2-simplices, where $0 \le w_j \le n$ controls the relative size among different $\Delta_j$.  Then the size of the 2-cycle $\tilde{\alpha}_2$ is $|\tilde{\alpha}_2|= \sum_j w_j$.   Now we further fine-grain each coarse simplex by increasing the number of vertices in it by a factor of $\beta$, i.e., there are $\beta n$ new vertices in each $\Delta_j$.  Due to the locally Euclidean feature, the fine graining leads to $O(\beta^{2/5} w_j)$ 2-simplices being intersected by $\tilde{\alpha}_2$ in each $\Delta_j$, and hence we get the size of the minimum 2-cycle as $|\tilde{\alpha}_2|=O(\beta^{2/5})  \sum_j w_j$.  We hence know that when increasing $m$, the 2-systole scales as $sys_2(M^5; \ZZ_2)=O(n^{2/5})$.
We can also get $sys_3(M^5; \ZZ_2)=O(n^{3/5})$ in a similar manner. We hence get the code distance as
\begin{equation}
d=O(n^{2/5}). 
\end{equation}
This family of 5D non-Abelian codes  hence have the code parameters $[[O(n), 3, O(n^{2/5})]]$.

Notably, this $O(n^{2/5})$ distance scaling  surpasses the $O(n^{1/3})$ distance barrier of transversal non-Clifford gate in conventional topological stabilizer code as implied by the Bravyi-K\"onig bound \cite{Bravyi:2013dx}, including  the 3D color/surface codes \cite{Bombin:2015jk, Kubica:2015, vasmer2019}  and the 6D self-correcting color code \cite{Bombin:2013cv} with transversal CCZ or $T$ gates. 

We also note that based on the transversal non-Clifford gates in the 3D color/surface code, a just-in-time-decoding (JIT) scheme was proposed to simulate the 3D code with a (2+1)D space-time, which requires only a 2D setup for the hardware \cite{Bombin:2018wj, Browneaay4929}.  However, this scheme is only trading time for space since the single-shot non-Clifford gate in the 3D code has to be implemented now for $O(d)$ time.  There is hence no reduction in the $O(d^3)$ space-time overhead.  In fact, a recent paper \cite{Davydova:2025ylx} has shown the equivalence for such a JIT decoding scheme to a non-Abelian $D_8$ quantum double model. 
One way to understand this is to use the twisted compatification scheme in Section \ref{sec:compactify} and Fig.~\ref{fig:compactification} for the specific case of $D=2$.   On the other hand, although our 5D non-Abelian self-correcting code can be obtained from the twisted compactification of the 6D self-correcting color code,  we directly find a single-shot non-Clifford logical gate acting on the compactified 5D code  without the need of an $O(d)$ time overhead.  Therefore, the total space-time overhead of our scheme is only $O(d^{5/2})$ instead of $O(d^3)$ in all the previous cases for topological codes.  This saving is mainly due to the use of higher-cup product which allows a cohomology operation of three 2-cocycles in five spatial dimension, which is impossible for a standard triple cup product \cite{zhu2023non, Hsin2024:classifying, zhu2025topological}.

\subsection{Universal logical gate set for the self-correcting quantum computer}

Although our present paper focuses on the logical non-Clifford gate scheme, here we also envision how it could be extended to a scheme for achieving a universal gate set.
Due to the Eastin-Knill no-go theorem \cite{Eastin:2009cj}, we do not expect that the 5D non-Abelian self-correcting code can achieve a universal gate set with transversal-like gates.   On the other hand, a 5D self-correcting Abelian loop (2-form) toric code can potentially achieve all the logical Clifford gates through generalized version of lattice surgery~\cite{doi:10.1126/sciadv.abn1717,Williamson:2024sky}.

Therefore, one can perform a code switching from the 5D Abelian 2-form toric code to the 5D non-Abelian self-correcting code  to perform the additional logical non-Clifford to achieve universality.  Such a fault-tolerant code switching via gauging and anyon condensation has been realized in Ref.~\cite{Davydova:2025ylx} in the 2D case, between twisted and untwisted quantum double.
Now due to the self-correcting properties of 5D codes, we expect that such a code switching can also be done in a single shot.  Therefore, one can potentially achieve a single-shot universal gate set with only $O(d^{5/2})$ space overhead, which could outperform either the scheme swtiching between 2D and 3D color codes \cite{Bombin:2015jk} or the 6D self-correcting color code scheme \cite{Bombin:2013cv} with $O(d^3)$ space overhead.

We also leave the discussion of the logical Clifford gates in the 5D non-Abelian self-correcting memory to Appendix \ref{sec:5D_Clifford}.

\section{Self-Correcting Quantum Memory}
\label{sec:self-correctingquantummemory}

\subsection{Interaction between system and environment}
\label{subsec:systembath}

To describe the self correcting quantum memory, we will model the interaction between the system and environment using Lindbladian evolution following e.g. \cite{Bombin_2013}. 
We will assume the total Hilbert space factorizes into the tensor product of the system Hilbert space and the environment Hilbert space.
The total system including the environment (``the bath") has the Hamiltonian
\begin{equation}
    H=H_\text{sys}+H_\text{env}+H_\text{int}~,
\end{equation}
where $H_\text{sys},H_\text{env},H_\text{int}$ are the Hamiltonians for the system, the environment, and interaction, respectively. The interaction Hamiltonian takes the form
\begin{equation}
    H_\text{int}=\sum_\alpha S_\alpha \otimes f_\alpha~,
\end{equation}
where $S_\alpha, f_\alpha$ act on the system Hilbert space and environment Hilbert space, respectively.
We can evolve a general operator using the total Hamiltonian and then trace out the environment Hilbert space. This results in effective Lindbladian evolution for an operator $O$ in Heisenberg picture: 
\cite{Bombin_2013}
\begin{align}
    \frac{dO}{dt}&={\cal L}[O],\cr
    {\cal L}&=i[H_\text{sys},O]+\frac{1}{2}\sum_\alpha\sum_\omega\left( h_\alpha(\omega)
    \left(S_\alpha^\dag(\omega)[O,S_\alpha(\omega)]\right.\right.\cr
    &\left.\left.\qquad\qquad\qquad \quad \quad +
    [S^\dag(\omega),O]S_\alpha(\omega)
    \right)\right)~,
\end{align}
where $S_\alpha(\omega)$ is the Fourier transform of $S_\alpha(t):=e^{iH_\text{sys}}S_\alpha e^{-iH_\text{sys}}$ , and $h_\alpha(\omega)$ is the Fourier transform of the autocorrelator of $f_\alpha$. We will assume $h_\text{max}:=\text{sup}_{\alpha,\omega\geq 0}h_\alpha(\omega)$ is finite \cite{Bombin_2013}.

From \cite{Bombin_2013}, the decay rate of a logical operator $O$ in the Heisenberg picture under the system Hamiltonian dressed by the error-correcting map onto the ground state is defined as
\begin{equation}
 -\langle O,{\cal L}(O)\rangle_\beta:=-\text{Tr}\left(\rho_\beta O^\dag {\cal L}(O)\right)~,
\end{equation}
where $\rho_\beta$ is the thermal equilibrium density matrix at temperature $1/(\beta k_B)$. Provided that $-\langle O,{\cal L}(O)\rangle_\beta \le \epsilon$, the decay of the operator is upper bounded as
\begin{align}
    \langle O, O(t)\rangle_\beta \ge e^{-\epsilon t}.
\end{align}

As shown in \cite{Bombin_2013}, for finite $h_\text{max}$ the upper bound of the decay rate $\epsilon$ can be given by the Boltzmann probability of the error syndrome, i.e. excitation pattern, that is critical for the logical operator $O$, i.e. single-qubit error that changes the value of $O$ (see discussions around Eq.~\eqref{eq:bound critical} for details)
\begin{equation}
 P_{\mathrm{crit}_O}=    \frac{\sum_\text{critical}e^{-\beta E}}{\sum_\text{all error} e^{-\beta E(\text{error})}}~.
\end{equation}
For a given temperature, if the above Boltzmann probability is exponentially small on the size of the system, we say that the logical operator is stable.
The discussion can be separated into two steps: \cite{Bombin_2013} (1) show that with high probability only configurations for small excitations appear in the Gibbs state. This is determined by the competition between the Boltzmann weight, and the entropy effect that counts the number of excitation patters with the same energy.
(2) Show that the configurations with only small (contractible) excitations are not critical, i.e. a single spin flip will not change the value of dressed observable. 
The second step essentially means that small excitations whose size is less than the code distance can be corrected, while the first step is more nontrivial and it gives the critical temperature for the quantum memory. We will investigate it in more details in Section \ref{sec:selfcorrecting}.

\subsubsection{Pauli noise model}

Let us first illustrate the above formalism using concrete Pauli noise model, following \cite{Bombin_2013}. The Pauli noise model is widely used and can describe practical noises in quantum computers \cite{Wallman:2015uzh}.
 Consider the Cubic theory Hamiltonian coupled to the environment,
\begin{equation}
    H=H_\text{sys}+H_\text{env}+H_\text{int}~,
\end{equation}
where
\begin{align}
    H_{\mathrm{sys}} = H'_{\text{Cubic}}~.
\end{align}

We consider the coupling to the environment in the form of
\begin{align}
    H_{\text{int}} = \sum_{j,k} X^j_k \otimes f_{j,k} + \sum_{j,k} Y^j_k \otimes g_{j,k} + \sum_{j,k} Z^j_k \otimes h_{j,k}~,
\end{align}
where the sum over $j$ runs $j=1,2,3$ and $k$ runs over all sites.
For simplicity, it is written as
\begin{align}
    H_{\text{int}} = \sum_{\sigma} \sigma\otimes f_\sigma~,
\end{align}
where the sum over $\sigma$ runs over all Pauli $X^j,Y^j,Z^j$ matrices on a single site.

The Cubic theory Hamiltonian with the above coupling is subject to Pauli errors. The $X$-error leads to the error syndrome of $H_{\text{Flux}
}$, characterized by $\pm 1$ eigenvalues of the terms of $H_{\text{Flux}
}$. These eigenvalues are represented by the vector $\mathbf{b}^X\in \Z_2^{3|c|}$.

The $Z$-error causes the syndrome of $H'_{\text{Gauss}}$. Due to the presence of the projector onto the zero flux sector, the eigenvalues of the terms in $H'_{\text{Gauss}}$ becomes either $1,0,-1$. In order to express the error syndrome in terms of the $\Z_2$ vector as in the case of the $Z$-error, one first needs to correct the $X$-errors by acting suitable Pauli $X$ operators. After the correction we have $\mathbf{b}^X=0$. The $Z$-error is then characterized by the vector $\mathbf{b}^Z \in \Z_2^{3|e|}$, which is the $\pm 1$ eigenvalues of the terms of $H'_{\text{Gauss}}$. 

Once the above sequential procedure of error correction is understood, the error syndrome of this model can be expressed by a pair of vectors
    \begin{align}
\mathbf{b} = \{\mathbf{b}^X, 
    \mathbf{b}^Z \}~,
\end{align}
where $\mathbf{b}^X\in \Z_2^{3|c|}$.

For a given stabilizer state in the presence of the errors $\mathbf{b}$, we define a recovery map that corrects the error, which is in the form of
\begin{align}
        \mathrm{corr}(\mathbf{b}) = \mathrm{corr}_\text{Gauss}(\mathbf{b}^Z) \mathrm{corr}_\text{Flux}(\mathbf{b}^X) ~.
    \end{align}
This map corrects the $X$-error first, and then $Z$-error next. Concretely,
\begin{align}
    \mathrm{corr}_\text{Flux}: \Z_2^{3|c|}\to \mathcal{P}_X~,
\end{align}
where $\mathrm{corr}_\text{Flux}(\mathbf{b}^X)$ is a product of $X^1,X^2,X^3$ that recovers the syndrome of the $H_{\text{Flux}}$. Also
\begin{align}
    \mathrm{corr}_\text{Gauss}: \Z_2^{3|e|}\to \mathcal{P}_Z~,
\end{align}
where $\mathrm{corr}_\text{Gauss}(\mathbf{b}^Z)$ is a product of $Z^1,Z^2,Z^3$ that recovers the syndrome of the $H_{\text{Gauss}}$. 

In the presence of the error, we consider the dressed logical operators
\begin{align}
    O_{\text{dr}} = \sum_{\mathbf{b} } \mathrm{corr}(\mathbf{b})^\dagger O \mathrm{corr}(\mathbf{b}) P_{\mathbf{b}}~,
\end{align}
where $O$ is the logical operator of the stabilizer state without errors. $P_{\mathbf{b}}$ is the projection onto the stabilizer state in the presence of the error $\mathbf{b}$.
The thermal stability of the quantum memory is verified by checking if the decay rate of the dressed logical operators $O_{\text{dr}}$ becomes exponentially small with respect to the system size, at nonzero low temperature.

For any Pauli operator $E\in\mathcal{P}$ and an error syndrome $\mathbf{b}$, the commutation relation between $O_{\text{dr}}$ and $E$ becomes a number
\begin{align}
    O_{\text{dr}} E P_{\mathbf{b}} = S(O,E,\mathbf{b}) E O_{\text{dr}}P_{\mathbf{b}}~,
    \label{eq:NE=SEN}
\end{align}
where $S(O,E,\mathbf{b})=\pm 1$. If there exists a single Pauli error $\sigma$ satisfying $S(O,E,\mathbf{b})=1$, the error $\mathbf{b}$ is said to be $O$-critical. That is, the logical operator $O_{\text{dr}}$ in the presence of the $O$-critical error changes its value by an additional single-qubit error.

Let us then obtain the upper bound on the decay rate. We use the inequality
\begin{align}
    -\langle A, \mathcal{L}(A)\rangle_\beta \le 2\hat{h}_{\text{max}} \sum_\sigma \langle[\sigma, A],[\sigma, A] \rangle_\beta~.
    \label{eq:boundbyhmax}
\end{align}
Using \eqref{eq:NE=SEN}, \eqref{eq:boundbyhmax} one can derive the upper bound of the decay rate as
\begin{align}
\begin{split}
    -\langle O_{\text{dr}}, \mathcal{L}(O_{\text{dr}})\rangle_\beta & \le 4\hat{h}_{\text{max}} \frac{\sum_{\mathbf{b},\sigma}e^{-\beta E_{\mathbf{b}}} (1-S(O,\sigma,\mathbf{b})) }{\sum_{\mathbf{b}}e^{-\beta E_{\mathbf{b}}} } \\
    & \le 8 \hat{h}_{\text{max}} n P_{\mathrm{crit}_O}~.
    \label{eq:bound critical}
\end{split}
\end{align}
Here, $P_{\mathrm{crit}_O}$ is the probability of finding the $O$-critical errors, i.e. the errors $\mathbf{b}$ where one can find a single qubit error $\sigma$ such that $S(O,\sigma,\mathbf{b})=-1$. $n$ is the number of physical qubits.

\subsubsection{Stability against small excitations}

As mentioned in Section \ref{subsec:systembath}, to achieve thermal stability of the quantum memory it is required that the small excitations are not critical for any logical operators. This can be satisfied by making an appropriate choice of the correction maps $\text{corr}(\mathbf{b})$ so that the small syndromes (say within the region $R$ of the size $\xi L$ for the system size $L$ and a constant $\xi$) disconnected from other syndromes 
is corrected by the Pauli operators within the region $R$.

\subsection{Self-correcting quantum memory with loops and membranes}
\label{sec:selfcorrecting}

Let us analyze the self-correcting property for the model on 5-dimensional hypercubic spatial lattice. The Hamiltonian model in Section \ref{sec:stringmembranes} with $l=m=n=2,D=6$ has Gauss law term for each edge and flux term for each cube, and each face has 3 types of qubits for $a,b,c$.

Since we have a commuting (non-Pauli) stabilizer models in Eq.~\eqref{eq:commuting_stabilizer}, the errors inducing the loop and membrane excitations can be decoupled and corrected independently, similar to the 4d self-correcting toric-code case where the e-excitations induced by the Z-type errors and m-excitations induced by the X-type errors can be corrected independently. Thus, we will apply a similar analysis using a Peierls argument as in Ref.~\cite{Dennis:2001nw} for the 4d toric code.  

We consider the quantum memory being coupled to a thermal bath with temperature $T$ and focus on the thermal noise since its not correctable for usual topological memories without self-correction capability.  

\paragraph{Errors inducing loop excitations}

We first discuss correction of the errors inducing loop excitations.   The unit-time probability for creation or survival of a small loop excitation on a plaquette, i.e. violating the Gauss term on the 4 boundary edges, is given by the Boltzmann ratio:
\begin{equation}\label{eq:rate1}
\frac{P(0\rightarrow 4)}{P(0\rightarrow 0)}= \frac{P(4\rightarrow 4)}{P(4\rightarrow 0)}= e^{-8\beta } \qquad (\beta=1/T), 
\end{equation}
where the energy cost along each edge is taken to be 2 (flipping $-1$ to $+1$ in the Gauss law term) and $P(i\rightarrow j)$ stands for the unit-time transition probability (rate) from $i$ violated edges to $j$ violated edges. Note that both transition processes $0\rightarrow 4$ and  $4\rightarrow 0$ can be induced by a ``plaquette flipping" corresponding to a Pauli-Z operator $Z_f$ acted on the qubit associated with the plaquette $f$. Similarly, the transition process between the configurations with 1 violated edge and 3 violated edges is also induced by a plaquette flipping, with the transition probability ratio given by:
\begin{equation}\label{eq:rate2}
\frac{P(1\rightarrow 3)}{P(1\rightarrow 1)}= \frac{P(3\rightarrow 3)}{P(3\rightarrow 1)}= e^{-4\beta }. 
\end{equation}
Finally, for all the configurations with two violated edges which have the same energy, one can set the plaquette flipping probability at unit time  to be $1/2$ which connects different  configurations and ensures ergodicity (any initial configuration has a non-zero probability to reach any final configuration).    As one can see, when the temperature $T$ (and the thermal noise) is low enough, such local system-bath interaction will gradually shrink the perimeter of the loop excitations and suppress the error proliferation autonomously.  This prevents a homologically non-trivial worldsheets of loop excitations being created which induces a logical error. 
 
The critical temperature below which the loop excitation error is self-correcting is determined by the balance between the Boltzmann suppression $e^{-2\beta\ell}$ for loop of length $\ell$ and the loop entropy.
The latter can be estimated from the self-avoiding random walk abundance $n(\ell)\sim P(\ell)\mu^\ell$, where $P(\ell)$ is a polynomial and $\mu$ is the connective constant on hypercubic lattice which is $\mu\sim 8.84$ for 5-dimensional hypercubic lattice \cite{madras2012self}.
Thus the critical temperature below which the loop error is self correcting is given by
\begin{equation}
    e^{-2\beta_c }\mu\geq  1\Rightarrow T_c\geq \frac{2}{\log \mu k_B}\sim 0.92/k_B~,
\end{equation}
where $\beta_c=1/T_c$, $k_B$ is the Boltzmann constant, and the energy is measured in the unit such that the smallest loop of four lattice edges has energy $2\times 4=8$. The above analysis is for one type of loop excitation, but it applies to three types of loop excitations, where the probability combining the Boltzmann factor and the entropy is raised to the third power. More precisely,
this means that below the critical temperature, there exists $L'> \xi L$ such that 
the critical probability for the logical membrane operator for loop excitations satisfy
\begin{align}\label{eqn:loopbound}
&P_\text{crit}<    \sum_{\ell>L'} n(\ell)e^{-2\beta \ell}\cr
&<\sum_{\ell>L'} e^{-(2\beta-\log\mu) \ell}\text{Poly}(\ell)
<\text{Poly}(L)\sum_{\ell>L'} e^{-(2\beta-\log\mu) \ell}
\cr
&<\text{Poly}(L)e^{-(2\beta-\log\mu) \xi L}\frac{1}{1-e^{-(2\beta-\log\mu)}}
\end{align}
is exponentially small in the system size, where we use $\text{Poly}(\ell)< \text{Poly}(L)$. 

We note that the above analysis also equivalently establishes a probabilistic local cellular-automaton decoder for the loop excitation errors in the context of active error correction, where the local decoder simulates the local system-bath interaction.  In particular, Eq.~\eqref{eq:rate1} and \eqref{eq:rate2}, along with the $1/2$-probability plaquette flipping for the configurations with two violated edges, define a local update rule of for the cellular-automaton decoder at each syndrome measurement cycle. Note that either the measurement or the recovery operation has certain error probability. An error threshold for this decoder can be obtained via the effective critical temperature: $p_c \sim e^{-8\beta_c}$ \cite{Dennis:2001nw}. Due to the local nature of the decoder, there is no non-local classical communication required in the decoding process, which hence does not lead to a classical communication/computation overhead as the system scales, in contrast to the case of the usual non-local decoders.

\paragraph{Errors inducing membrane excitations}

Now let us analyze the membrane excitation error in a similar manner. The smallest membrane excitation corresponds to violating the flux term on a cube $s_3$, which is a $5-3=2$ dimensional membrane on the dual lattice. Let us consider the flux for $a$. Then the Gauss law term for $b,c$ will have energy cost 1 for the edge $s_1,s_1'$ such that $\int \tilde s_3\cup \tilde s_1\cup \tilde s_1'=1$.
The cube $s_3$ corresponds to one surface $s_2$ such that $\int \tilde s_3\cup \tilde s_2=1$ on 5-dimensional hypercubic lattice, and the edges $s_1,s_1'$ can be two ordered consecutive edges on the boundary of the surface $s_2$. Thus there are four Gauss law terms contribute to energy cost $1\times 4=4$, with a total energy $4\times 1+2\times 1=6$, where the second term is from the flux term on the cube. 
The smallest membrane excitation created by a single Pauli $X^i_f$, say $i=1$, consists of 6 cubes from the $2\times (5-2)=6$ directions perpendicular to the face $f$ in both positive and negative directions. 
Let us consider the following transition probabilities:
\begin{itemize}
    \item Creation or survival of a small membrane excitation, i.e. fluxes on 6 cubes:
\begin{equation}\label{eq:membrane_transition1}
    \frac{P(0\rightarrow 6)}{P(0\rightarrow 0)}=\frac{P(6\rightarrow 6)}{P(6\rightarrow 0)}=e^{-36\beta}~.
\end{equation}

    \item Deforming the membrane excitation with fluxes on $1\rightarrow 5$ cubes or $5\rightarrow 1$ cubes:
\begin{equation}\label{eq:membrane_transition2}
    \frac{P(1\rightarrow 5)}{P(1\rightarrow 1)}=\frac{P(5\rightarrow 5)}{P(5\rightarrow 1)}=e^{-24\beta}~.
\end{equation}

    \item Deforming the membrane excitation with fluxes on $2\rightarrow 4$ cubes or $4\rightarrow 2$ cubes:
\begin{equation}\label{eq:membrane_transition3}
    \frac{P(2\rightarrow 4)}{P(2\rightarrow 2)}=\frac{P(4\rightarrow 4)}{P(4\rightarrow 2)}=e^{-12\beta}~.
\end{equation}

    \item Deforming the membrane excitation with fluxes on $3\rightarrow 3$ cubes: one can set the cube flipping probability at unit time to be $1/2$ which connects different configurations and ensures ergodicity.
\end{itemize}

To estimate the critical temperature, we will need to balance the Boltzmann suppression $e^{-6\beta A}$ for membrane excitations of area $A$ and the entropy of the membrane.
The latter can be estimated from self-avoiding random surfaces on 5-dimensional hypercubic dual lattice, which is discussed in e.g.~Ref.~\cite{DURHUUS1983185}. The abundance again grows as $e^{A/\tau_0}$ for some constant $\tau_0$, thus the critical temperature is $T_c'\geq 6\tau_0/k_B$.
A probabilistic cellular automaton decoder is also present for the membrane excitation errors, similar to the one discussed above for the loop excitation errors.  
Below the critical temperature, there exists $A'> (\xi L)^2$ such that 
the critical probability for the logical volume operator for membrane excitations satisfy
\begin{align}\label{eqn:membranebound}
&P_\text{crit}<    \sum_{A>A'} n(A)e^{-6\beta A}\cr
&<\sum_{A>A'} e^{-(6\beta-1/\tau_0) A}\text{Poly}(A)
<\text{Poly}(L)\sum_{A>A'} e^{-(6\beta-1/\tau_0) \ell}
\cr
&<\text{Poly}(L)e^{-(6\beta-1/\tau_0) (\xi L)^2}\frac{1}{1-e^{-(6\beta-1/\tau_0)}}
\end{align}
is exponentially small in the system size, where we use $\text{Poly}(A)< \text{Poly}(L)$. 

Combining the results from the loop and membrane excitations, for temperature below min$(T_c,T_c')$,
using the equations (\ref{eq:bound critical}), (\ref{eqn:loopbound}) and (\ref{eqn:membranebound}) we complete the proof that the
 quantum memory is self-correcting.  

Note that the above transition probablities (Eq.~\eqref{eq:membrane_transition1} to \eqref{eq:membrane_transition3} and the additional 1/2-probability transition rule) also equivalently establish a probablistic local cellular automaton decoder for the membrane excitation errors.

\subsection{Readout of the memory and decoding the logical information}
We have analyzed the error suppression during the process of the quantum computation, while there always exist residual errors that are never completely corrected.   In the end of the quantum computation, we need to measure the encoded logical qubits in a certain logical basis. This can be done by performing a readout of all the qubits in a certain basis and then measure the corresponding logical operators.  A convenient choice for this code is to measure all the qubits in the $Z$ basis, which enables the measurement of electric Wilson membrane operators which have a product form $\prod Z^i_f$.   Now in order to extract the encoded logical information, we assume that a reliable classical computer can help with decoding the measured classical data.  Note that the measured data will contain errors originated from e.g.~the thermal and readout noise.  The errors in the Pauli-Z measurement can be detected via the error syndromes corresponding to the Gauss term, i.e., loop excitations.   We then apply the probabilistic local cellular automaton decoder (for the loop excitation errors) presented above to clean up the loop excitations. Since the classical computer is reliable, this recovery process can be executed perfectly. One can than perform a readout on the Wilson membrane operators $\prod Z^i_f$. Note that a logical error can occur if the above recovery process creates a homologically non-trivial worldsheet of the loop excitations.   Such a logical error can be exponentially suppressed with the system size $L$ when the bath temperature is below the critical temperature min$(T_c,T_c')$ and the readout error is below the error threshold of this decoder $p_c \sim e^{-8\beta_c}$.

\subsection{Generalization: general Cubic theory without particles}

Let us generalize the argument to general Cubic theory without particles, with $m,n,D$ satisfying (\ref{eqn:noparticles}).
 The theory consists of electric excitations of $(m-1),(n-1),(D-m-n-1)\geq 1$ dimensions and magnetic excitations of $(D-m-2),(D-n-2),(m+n-2)\geq 1$ dimensions.
The thermal stability depends on the balance between the Boltzmann suppression and the entropy.
\begin{itemize}
    \item 
The energy cost for the $k$-dimensional electric excitations is $2\epsilon_0 V_k$ where the Hamiltonian has coefficient $\epsilon_0$ and $V_k$ is the volume of the excitations in terms of the lattice spacing. 
The energy cost for the $k'$-dimensional magnetic excitations is $(\epsilon_0+2\epsilon_0)V_{k'}=3\epsilon_0 V_{k'}$.
The energy gap is thus bounded below by $2\epsilon_0 V_k$ for $k$-dimensional excitation.

\item 
The multiplicity of $k$-dimensional excitation of volume $V_k$ for $k\geq 1$ can be estimated as follows, following \cite{Dennis:2001nw}. The first step of the random ``submanifold-walk'' on hypercubic lattice chooses $k$ distinct directions out of $2D$ directions (including the orientation), and the next step on chooses a boundary of this $k$-simplex to attach with another $k$-simplex, thus one first make a choice out of $2k$ (including the orientation), and then one attaches to the original $k$-simplex with a new one by combining the boundary with one orthogonal direction not overlap with the original $k$-simplex, which is a choice of $2D-k$.
Thus the multiplicity of $V_k$ steps is bounded above by
\begin{equation}
    n(V_k)\leq 2D(2k(2D-k))^{V_k}~.
\end{equation}
\end{itemize}

To suppress large excitations, it is sufficient to require the temperature to be lower than $1/(\beta k_B)$ where
\begin{equation}
    e^{-\beta 2\epsilon_0 V_k} 2D(2k(2D-k))^{V_k}\sim 1~.
\end{equation}
In other words, this gives an estimation of the critical temperature below which there is no large $k$-dimensional excitations
\begin{equation}
    T_c(k)\sim \frac{2\epsilon_0}{\log (2k(2D-k))k_B}~.
\end{equation}
The critical temperature for all excitations is then 
\begin{align}
    T_c\sim & \text{min}_{k\in S_{m,n}}    T_c(k),\cr 
    S_{m,n}:=&\{(l-1),(m-1),(n-1),\cr &\; (D-l-2),(D-m-2),(D-n-2)\}
    ~.
\end{align}

We remark that if $k\rightarrow 0$, i.e. particle excitations, the estimation gives $T_c(k)\rightarrow  0$ and $T_c\rightarrow 0$, which is consistent with thermal instability for topological orders with particles.

\section{Discussion and Outlook}
\label{sec:discussion}

In this work we have constructed a family of infinite many candidate non-Abelian self-correcting quantum memories in spacetime dimension $D\geq 5+1$, which is the dimension that the non-Abelian self-correcting memories can occur.  These models also give rise to the first set of non-Abelian topological orders which are stable at finite temperature. The models can be described by three $\mathbb{Z}_2$ higher-form gauge fields with cubic topological interaction among them. In these models, the electric excitations obey Abelian fusion rules, while the magnetic excitations obey non-Abelian fusion rules. We discuss the properties of the models using field theory and non-Pauli stabilizer lattice Hamiltonian models. We show that the non-Abelian self-correcting memory in 5D has a nearly transversal non-Clifford CCZ gate with the distance scaling $d=O(n^{2/5})$ that overcomes the $n^{1/3}$ distance barrier in the color codes.
We rigorously prove the self-correcting properties  with a lower bound of the memory time and the thermal stability, and devise a probabilistic local cellular-automaton decoder for these codes.

There are several future directions. One can develop a deterministic cellular-automaton decoder using non-equilibrium dynamics in analogy to Toom's rule or sweep decoder \cite{dua2023quantum},  which is more efficient in cleaning up the errors than the current probabilistic decoder. 

One can further extend the current non-Clifford gate scheme in the 5D non-Abelian self-correcting code to a constant-time full universal gate set through single-shot code switching between the 4D loop toric codes which can admit a full Clifford group and the 5D non-Abelian code which admits the non-Clifford logical CCZ gate.

The logical gates in our models can also be constructed using the gauging and anyon condensation methods in \cite{Huang:2025cvt,Davydova:2025ylx}. We will present the constructions in future work.

Finally, in our discussion we have focused on topological orders with fully-mobile excitations. It would be interesting to explore more general fractonic non-Abelian quantum memories in even lower spatial dimensions, where the code space can have exponentially many logical qubits in the system size.

\section*{Acknowledgments}

We thank Yu-An Chen and Nathanan Tantivasadakarn for discussions. We thank Maissam Barkeshli, Xie Chen, Meng Cheng, Shu-Heng Shao and Dominic J. Williamson for comments on a draft.
P.-S.H. was supported by Simons
Collaboration of Global Categorical Symmetry, Department of Mathematics King’s College
London, and also supported in part by grant NSF PHY-2309135 to the Kavli Institute for
Theoretical Physics (KITP). 
R.K. is supported by the JQI postdoctoral fellowship at the University of Maryland, U.S. Department of Energy through grant number DE-SC0009988 and the Sivian Fund.  G.Z. is supported by the U.S. Department of Energy, Office of Science, National Quantum
Information Science Research Centers, Co-design Center
for Quantum Advantage (C2QA) under contract number
DE-SC0012704.
P-S.H. thanks Kavli Institute for Theoretical Physics for hosting
the program “Correlated Gapless Quantum Matter” in 2024, during which part of the work
is completed. We thank Peter Shor for posting the question about non-Abelian topological orders on Twitter, which partially inspired this work.

\appendix

\section{Review of Cup Product on Triangulated and Hypercubic Lattices}
\label{sec:cupproductrev}

Let us review cup product for $\mathbb{Z}_2$ valued cochains on triangulated lattice and hypercubic lattice. For more details, see e.g. \cite{milnor1974characteristic,Benini:2018reh, PhysRevB.101.035101,Chen:2021ppt,Chen:2021xuc}.

A $\mathbb{Z}_2$-valued $m$-cochain $\alpha_m$ is a map from $m$-simplices on the lattice to $0,1$ mod 2.
The cup product of $m$-cochain $\alpha_m$ and $n$-cochain $\beta_n$ is an $(m+n)$-cochain $\alpha_m\cup \beta_n$ defined as follows:
\begin{itemize}
    \item For triangulated lattice, denote $k$-simplices by the vertices $s_k=(0,1,2,3,\cdots, k)$, the cup product takes the following value on $(m+n)$-simplices:
    \begin{align}
        &\alpha_m\cup \beta_n(0,1,2,\cdots, m+n)\cr
        &\; =
        \alpha_m(0,1,\cdots,m)\beta_n(m,m+1,\cdots,m+n)~.
    \end{align}
    We note that there is a common vertex $m$.

    \item For hypercubic lattice, the cup product $\alpha_m\cup \beta_n$ on $(m+n)$-dimensional hypercube $s_{m+n}$ that span the coordinates $(x^1,x^2,\cdots x^{m+n})\in [0,1]^{m+n}$ is given by
    \begin{align}
        &\alpha_m\cup \beta_n(s_{m+n})\cr
        & =\sum_I \alpha_m\left([0,1]^I\right)\beta_n\left((x^I=1,x^{\bar I}=0)+[0,1]^{\bar I}\right)~,
    \end{align}
    where the summation is over the different sets $I$ of $m$ coordinates out of the $(m+n)$ coordinates $x^1,\cdots x^{m+n}$, $\bar I$ denotes the remaining $n$ coordinates. 
    In each term in the sum, $\alpha_m,\beta_n$ are evaluated on an $m$-dimensional hypercube and an $n$-dimensional hypercube, where the two hypercubes intersect at a point $(x^I=1,x^{\bar I}=0)$:
    \begin{itemize}
        \item $[0,1]^I$ is the $m$-dimensional unit hypercube starting from $(x^I=0,x^{\bar I}=0)$ and ending at $(x^I=1,x^{\bar I}=0)$, i.e. $[0,1]^I=\{0\leq x^i\leq 1,x^j=0:i\in I, j\in \bar I\}$.
        \item     $(x^I=1,x^{\bar I}=0)+[0,1]^{\bar I}$ is the $n$-dimensional hypercube in the $\bar I$ directions starting from $(x^I=1,x^{\bar I}=0)$ and ending at $(x^I=1,x^{\bar I}=1)$, i.e. $(x^I=1,x^{\bar I}=0)+[0,1]^{\bar I}=\{x^i=1,0\leq x^j\leq 1:i\in I,j\in \bar I\}$.
    \end{itemize}

\end{itemize}

\subsection{Higher cup product}
\label{sec:highercup}

\begin{figure}[t]
    \centering
    \includegraphics[width=0.7\linewidth]{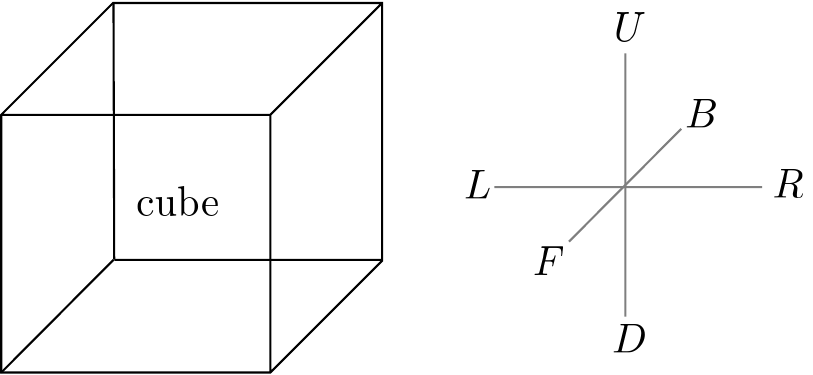}
    \caption{Faces of a cube are labeled by $L,R,B,F,U,D$.}
    \label{fig:cube_labels}
\end{figure}

There are generalizations of cup product called $\cup_i$ product due to Steenrod \cite{Steenrod1947}, with $\cup_0=\cup$ being the ordinary cup product. In our discussion we only use $\alpha\cup_1 \beta$ for 2-cochains $\alpha,\beta$.
This is a $2+2-1=3$ cochain, whose value on 3-cell is as follows:
\begin{itemize}
    \item For triangulated lattice, denote the 3-simplex by its vertices $(0123)$, the value of $\alpha\cup_1 \beta$ is
    \begin{equation}
        \alpha\cup_1 \beta (0123)=\alpha(023)\beta(012)+\alpha(013)\beta(123)~.
    \end{equation}

    \item For hypercubic lattice, the value of $\alpha\cup_1\beta$ on a 3-dimensional cube shown in Fig.~\ref{fig:cube_labels} is
    \begin{align}
        \alpha\cup_1\beta(\text{cube})
        &=
        \alpha(L)\beta(B)+\alpha(L)\beta(D)\cr 
        &\quad +\alpha(B)\beta(D)+\alpha(U)\beta(F)\cr
        &\quad +\alpha(U)\beta(R)+\alpha(F)\beta(R)~,
    \end{align}
    where $L,R$ are the left and right side faces of the cube, $B,F$ are the back and front faces of the cube, and $U,D$ are the up and down faces of the cube. 
    
\end{itemize}
Note that the expression of the higher cup product $\cup_1$ involves 2-cells that overlap along 1-cell. 

 \section{Logical CZ, S, Hadamard Gates in 4D Loop Toric Code}
\label{sec:loopTC}

We will discuss the logical gates in the loop toric code model  in (4+1)D, which is an example of self-correcting quantum memory at finite temperature (e.g. \cite{Dennis:2001nw,Brown_2016}).
The Hamiltonian is a stabilizer model:
\begin{equation}
    H=H_\text{Gauss}+H_\text{Flux}=-\sum_e \prod_{e\in \partial f} X_f-\sum_c \prod_{f\in\partial c} Z_f~,
\end{equation}
where the $\mathbb{Z}_2$ 2-form gauge field is $a_f=(1-Z_f)/2$.
The ground states on $M_4$ 4-manifold is given by $q$ qubits with $H^2(M_4,\mathbb{Z}_2)=\mathbb{Z}_2^q$. 

The loop toric code has loop excitations labelled by the electric and magnetic charges $(q_e,q_m)$ with $q_e,q_m=0,1$, i.e. pure electric loop $(1,0)$, pure magnetic loop $(0,1)$ and dyon loop $(1,1)$, corresponding to violation of the Gauss law terms along the loop, flux terms along the loop and both types of Hamiltonian terms along the loop.

\subsection{Hadamard gate}

The model on hypercubic lattice has electromagnetic duality spin rotation symmetry that exchanges $X\leftrightarrow Z$, and swaps the lattice with the dual lattice.
To see this, we note that the Gauss law term on edge in the $x$ direction has product over $X_f$ spanning $x$ direction and one of the remaining $4-1=3$ directions, and taking into account the orientation the product is over $3\times 2=6$ $X_f$. Similarly, each cube $c$ spanning $x,y,z$ directions has faces in $xy,yz,zx$ directions, with orientations there are in total $3\times 2=6$ $Z_f$ in the product of each flux term.

On the loop excitations the symmetry acts as
$S:(q_e,q_m)\leftrightarrow (q_m,q_e)$, where we note $-q_m=q_m$ mod 2, as well as $T:(q_e,q_m)\rightarrow (q_e+q_m,q_m)$ \cite{Gaiotto:2014kfa,Chen:2021xuc}. The $S$ symmetry acts as Hadamard gate.

\subsection{CZ and S gates}

The $T:(q_e,q_m)\rightarrow (q_e+q_m,q_m)$ symmetry is generated by the gauged SPT defect decorated with $H^4(B^2\mathbb{Z}_2,U(1))=\mathbb{Z}_4$: \cite{Chen:2021xuc}
\begin{equation}
    U(M_4)=i^{\int {\cal P}(a)}~.
\end{equation}
The gauged SPT phase on the domain wall is the semion Walker Wang model.
The operator acting the ground states labelled by holonomy $\{n_i=0,1\}$ for $i=1,2,\cdots, q$ according to the intersection form $M_{ij}$ on $H^2(M_4)$:
\begin{equation}
    U(M_4)|\{n_i\}\rangle=i^{\sum_j M_{jj}n_j^2} (-1)^{\sum_{i<j}M_{ij}n_in_j}|\{n_i\}\rangle~.
\end{equation}

Consider the following cases:
\begin{itemize}
    \item $M_4=T^4_{x,y,z,w}$, with $q=6$. The operator realizes the logical gate
    \begin{equation}
    U(T^4)=\text{CZ}_{xy,zw}\text{CZ}_{xz,yw}\text{CZ}_{yz,xw}~,
    \end{equation}
    where $xy$ is the 2-cycle spanning $x,y$ directions, and similarly for $zw,xz,yw,yz,xw$.

In particular, this implies that $U(T^4)X_{xy}U(T^4)^{-1}=X_{xy}Z_{zw}$, where $X,Z$ are the logical gates. Thus the symmetry permutes the magnetic loop excitation to the dyon loop excitation. This is the case considered in \cite{Chen:2021xuc}.

    \item $M_4=\mathbb{CP}^2$, with $q=1$.
    The operator realizes the logical gate
    \begin{equation}
    U(T^4)=S~.
    \end{equation}
    In other words, if the state has trivial holonomy on the 2-cycle, the operator acts trivially; if the state has nontrivial holonomy on the 2-cycle, the operator acts as $i$.

\end{itemize}

\section{Logical Clifford Gates in the 5D Non-Abelian Self-correcting Code}\label{sec:5D_Clifford}
In addition to the non-Clifford CCZ logical gate, we can also do some of logical Clifford gates using constant depth circuits following \cite{Hsin:2024nwc} on space of other topologies.

For example, take the space to have topology $\mathbb{CP}^2\times S^1$, and focus on the logical qubit $a=nw_2$ for $w_2$ of $\mathbb{CP}^2$.
The 1-form symmetry operator $i^{\int {\cal P}(a)}$ gives addressable logical S gate:
\begin{equation}
    i^{\int_{\mathbb{CP}^2} {\cal P}(a)}=i^{[n]}=\overline{S}~,
\end{equation}
where $\mathcal{P}(b)$ is the Pontryagin square, 
\begin{align}
    \mathcal{P}(b) = b\cup b - b\cup_1 db~,
\end{align}
which defines a $\Z_4$ valued 4-cocycle.
$[n]$ denotes the mod 2 restriction of $n$ to $0,1$.

Similarly, take the space to have topology $S^2\times S^2\times S^1$, and the logical qubit $a=n_1 \kappa_1+n_2\kappa_2$ for $\kappa_1,\kappa_2$ the K\"ahler forms of the two $S^2$s. The 1-form symmetry operator
$i^{\int {\cal P}(a)}$ gives addressable logical CZ gate:
\begin{equation}
    i^{\int_{S^2\times S^2} {\cal P}(a)}=(-1)^{n_1n_2}=\overline{CZ}_{n_1,n_2}~.
\end{equation}

 \bibliography{biblio, mybib_merge}
 \bibliographystyle{apsrev4-2}

\end{document}